\documentclass[conference]{IEEEtran}
\usepackage{cite}
\usepackage{amsmath,amssymb,amsfonts}
\usepackage{graphicx}
\usepackage{textcomp}
\usepackage{xcolor}
\usepackage[hyphens]{url}

\usepackage{float}
\usepackage{tcolorbox}
\usepackage{graphicx}
\usepackage{subcaption}
\usepackage{amsmath}
\usepackage{multirow}
\usepackage{algorithm}
\usepackage{algpseudocode}
\usepackage{tikz,xspace}

\newcommand{\methodname}{\emph{FaaSched}\xspace}
\newcommand*\circled[1]{\raisebox{.5pt}{\textcircled{\raisebox{-.9pt} {#1}}}}

\def\BibTeX{{\rm B\kern-.05em{\sc i\kern-.025em b}\kern-.08em
    T\kern-.1667em\lower.7ex\hbox{E}\kern-.125emX}}

\title{FaaSched: A Jitter-Aware Serverless Scheduler}
\author{\IEEEauthorblockN{Abhisek Panda}
	\IEEEauthorblockA{Department of Computer Science \\
		Indian Institute of Technology, Delhi\\
		New Delhi, India \\
		Email: abhisek.panda@cse.iitd.ac.in}
	\and
	\IEEEauthorblockN{Smruti R. Sarangi}
	\IEEEauthorblockA{Department of Computer Science \\
		Indian Institute of Technology, Delhi\\
		New Delhi, India \\
		Email: srsarangi@cse.iitd.ac.in}
}

\begin{document}
\maketitle
\thispagestyle{plain}
\pagestyle{plain}

\begin{abstract}

Serverless computing systems are becoming very popular. Large corporations such as Netflix, Airbnb, and Coca-Cola use such systems for running their websites and IT systems. The advantages of such systems include superior support for auto-scaling, load balancing, and fast distributed processing. These are multi-QoS systems where different classes of applications have different latency and jitter (variation in the latency) requirements: we consider a mix of latency-sensitive (LS) and latency-desirable (LD) applications. Ensuring proper schedulability and QoS enforcement of LS applications is non-trivial. We need to minimize the jitter without increasing the response latency of LS applications, and we also need to keep the degradation of the response latency of LD applications in check. 

This is the first paper in this domain that achieves a trade-off between the jitter suffered by LS applications and the response latency of LD applications. We minimize the former with a bound on the latter using a reinforcement learning (RL) based scheme. To design such an RL scheme, we performed detailed characterization studies to find the input variables of interest, defined novel state representations, and proposed a bespoke reward function that allows us to achieve this trade-off. For an aggressive use case comprising five popular LS and LD applications each, we show a reduction in response time variance and mean latency of 50.31\% and 27.4\%, respectively, for LS applications. The mean degradation in the execution latency of LD applications was limited to 19.88\%.

\end{abstract}

\section{Introduction}
\label{sec:intro}

Online applications are transforming their monolithic architectures into microservice-based architectures due to the latter's enhanced scalability and modularity~\cite{microservice1, microservice2, microservice3}. Thus, serverless computing has emerged as a popular cloud paradigm, where the execution is broken down into running a series of small code snippets ({\em functions}) across a set of distributed nodes~\cite{serverless_1}. Each function runs within a sandbox.  Over the last few years several large cloud computing vendors have adopted this paradigm and there are large platforms in place for running serverless applications such as Microsoft Azure Functions~\cite{azure}, Amazon Lambda~\cite{lambda}, Google Cloud Functions~\cite{google} and IBM Cloud Functions~\cite{ibm}. Such serverless platforms provide auto-scaling features and autonomously manage the infrastructure. A large number of applications now run on such systems~\cite{utilization,lass,ensure,sebs}. Such applications span a large number of domains: IoT, web services, security, machine learning, data analytics, scientific, and multimedia applications.

According to a recent survey, 62\% of serverless applications have stringent latency requirements~\cite{latency_req}. We can thus categorize serverless applications into two categories based on their latency requirements: latency sensitive (LS) and latency desirable (LD) (see Table~\ref{tab:category}) (similar to \cite{provisioned_concurrency}). LS applications have strict performance requirements, and even slight variations in the response latency can impact revenues~\cite{variation1, variation2,variation3,variation4}. The variation (referred to as \emph{jitter}) in the response latency of a serverless function makes it difficult to predict the end-to-end latency of a serverless application~\cite{cloud_jitter} and consequently makes it hard to satisfy strict quality-of-service requirements. This makes it hard to satisfy schedulability constraints of even soft real-time applications and guarantee practical upper bounds~\cite{goodspread,schedulability}.  On the other hand, LD applications can sustain a larger jitter. Keeping this in mind, Amazon charges an additional 20 cents for provisioned concurrency (greater control on the execution time and jitter, see~\cite{goodspread}).  In prior work, authors have identified variable initialization times~\cite{catalyzer,nightcore} and sharing of compute and memory resources~\cite{interference} as the causative factors of jitter.

\begin{table}
	\centering
	\footnotesize
    \caption{Categorization of serverless applications: (LS: Latency sensitive, LD: Latency Desirable)}
	\label{tab:category}
	\begin{tabular}{|p{1.1cm}|p{6cm}|}
		\hline
		\textbf{Category}	&	\textbf{Description}	\\
		\hline
        \textbf{LS}       & The latency of an application \emph{must} be lower than a pre-specified deadline. Examples:
        web search, e-commerce, security, IoT, and real-time collaboration applications.	\\
		\hline
        \textbf{LD}       & The jitter in execution times is not important as long as it stays within reasonable limits.
        Examples: data collection and visualization, video analytics, machine learning training, and other batch mode
        applications.	\\
		\hline
	\end{tabular}
\end{table}

To reduce jitter in the latency of an application, prior work focused on designing efficient resource scheduling schemes. Ensure~\cite{ensure} and the work by Kaffes et. al.~\cite{placement2} tried to optimize the allocated CPU time and the CPU core partitioning technique to mitigate resource contention. However, they do not consider the impact of resource allocation decisions on the other applications running in the system. Instead of allocating resources to a single application, LaSS~\cite{lass} uses a queuing theory-based model to spawn an ideal number of container processes for an application based on the arrival rate and execution latency (subject to simplistic assumptions regarding determinism). Sadly, the jitter in the execution latency caused by colocation is overlooked in the theoretical modeling part itself.

In the serverless computing paradigm, the number of containers corresponding to a function varies drastically with the service time and arrival rate of the function. Therefore, the degree of interference due to colocation is dependent on the aforementioned parameters. Prior work on QoS~\cite{parties, sinan, clite} has designed resource partitioning schemes for long-running microservices, which are not suitable for ephemeral and dynamic serverless workloads. Moreover, serverless providers are responsible for managing the allocated resources of their customers or clients. As it is a multi-tenant setup, they must ensure a higher degree of fairness among the tenants.

This motivated us to identify the primary sources of jitter in a scenario with application colocation. We performed a correlation study between the execution latency of a request and different SW and HW-level events. We found that the following broad factors contribute to jitter: \emph{CPU contention}, \emph{locking mechanisms} and \emph{code locality}. To the best of our knowledge, such a study and subsequent mitigation mechanisms based on these insights are new in the serverless community.

In terms of the knobs for jitter mitigation, we focus on the scheduling policy and core affinity. Tuning these knobs turned out to be quite difficult. The traditional adage that says, higher the priority better it is for the application, does not hold here. This is because this depresses the priority of OS threads and as a result system calls take longer (also observed in~\cite{jitter,jitter2}). Moreover, there could be other LS applications, and we would not like to degrade the performance of LD applications beyond a point. Moreover, if we couple these decisions with CPU affinity or in other words core partitioning decisions, then the optimization space becomes even more complex. We found that simple heuristics or queuing theory based models with simplistic assumptions about the environment do not work. There is a need to look at modern methods such as reinforcement learning (RL). We were further encouraged by prior work in virtual machine allocation in clouds~\cite{rl_cpu2,vconf}. Even though the problems are very different, however, RL-based techniques have shown a lot of promise and are thus worthy candidates for evaluation.

Our proposed \methodname uses an \emph{RL} scheme to set the two knobs -- real-time scheduling priority of each application and its CPU affinity. The key tasks are finding the right set of input features based on detailed correlation studies, choosing the right RL model, and designing appropriate reward functions. The last point was particularly vexing because we need to minimize the jitter of LS applications, preferably speed them up also, and keep the performance of degradation of LD applications in check (the {\em fairness} part). Moreover, when we have a large number of applications, we need to ensure that the number of dedicated cores is less than the total number of cores and the OS has enough breathing space (time and resources to execute system calls and perform bookkeeping). Note that our scheme is a pure software implementation and does not require any hardware support.

To summarize, our contributions are as follows:
\begin{enumerate}
    \item We identify the key sources of jitter in a system when multiple serverless applications are colocated on a single host. Our key finding is that CPU contention, locking mechanisms, and code locality have a significant impact on execution latency jitter.

    \item We design \methodname, which sets the priority and the physical CPU core allocation of LS applications while preventing these applications from monopolizing CPU resources.

    \item Our scheme improves the variance and the mean of the response latency of LS applications by 50.31\% and 27.4\%, respectively, over the current state of the art~\cite{lass}. We consider 5 LS applications and 5 LD applications, all running   simultaneously.

	\item The degradation in the mean execution latency of LD applications is limited to 19.88\%.
\end{enumerate}

The rest of the paper is organized as follows. We discuss the relevant background for the paper in Section~\ref{sec:background}. Subsequently, we profile microarchitectural counters and collect the execution statistics of an LS application when colocated with an LD application in Section~\ref{sec:characterization}. We motivate the need for an RL scheme to set the priority and the physical CPU core allocation of LS applications in Section~\ref{sec:motivation}. This is followed by the design of \methodname in Section~\ref{sec:design}. Section~\ref{sec:eval} evaluates the efficacy of \methodname for reducing jitter. We discuss the related work in Section~\ref{sec:related_work}, and finally, we conclude in Section~\ref{sec:conclusion}.

\section{Background}
\label{sec:background}
In this section, we provide the relevant background about serverless computing and reinforcement learning.

\subsection{Serverless Computing}
The serverless computing paradigm basically divides a large task into several smaller sub-tasks -- each one is known as a serverless {\em function} that runs in an isolated environment. These functions can run on different network nodes in a distributed fashion. This idea is basically a second avatar of erstwhile web service based architectures; in this case, we can operate at the level of functions because of ultra-fast network speeds. To improve performance, nodes can keep read-only data and code warm (in their caches). Let us elaborate.

On a serverless computing platform, a developer only needs to write an application in the form of a {\em function chain}. In response to an external event, the platform executes each function of the function chain in a sandbox. To restrict a user's monopoly on resources, the user has to define resource requirements at the granularity of functions. In contrast to conventional cloud computing, such platforms provide auto-scaling features. In addition, it is easy to perform load balancing because we can just bring up more nodes that host the code of a certain frequently-accessed function. This paradigm is especially suitable for latency-sensitive (LS) applications like web applications, security applications, and IoT applications because functions have dedicated resources such as machines, and it is possible to optimize their run time environment. 

Such an environment has three major attributes: the number of idle sandboxes (\emph{warm}) created for a function on the system; resource configuration, such as the number of sandboxes created for a function; and the degree of colocation of tasks that share resources.

\subsubsection{Auto-scaling Features within a Single Node}
A serverless platform spawns multiple sandbox processes for an application to meet its latency requirements. The number of sandbox processes associated with an application varies with the application's request arrival rate. In Apache Openwhisk~\cite{openwhisk}, each of the sandboxes of an application is mapped to three multi-threaded processes: \emph{containerd-shim, entrypoint script}, and \emph{web server}.

\subsubsection{Cold Start}
Due to the on-demand execution of requests and associated security considerations, the platform spawns an ephemeral sandbox to serve a request. Therefore, the process of spawning a sandbox is on the critical path -- this is referred to as \emph{cold start}. Cold start is a problem for LS applications when they have a high execution rate~\cite{nightcore}. To minimize the cold start latency, a serverless platform may use a light-weight sandbox mechanism~\cite{nightcore}, pre-warming, warm containers~\cite{warm_lambda, warm_whisk}, or checkpoint and restore-based techniques~\cite{catalyzer, prebaking, reap}. Nowadays, platforms use a model predictive control mechanism to predict the arrival rate of function invocations and spawn the required number of sandboxes in a cluster~\cite{pred1, pred2}.

\subsubsection{Multi-tenant Setup}
In a typical multi-tenant setup, the serverless platform hosts multiple functions with varying arrival rates on the same machine; they share the limited compute and memory resources of the host among themselves. Therefore, colocation introduces interference in the function execution and leads to jitter in the response latency~\cite{interference, faasrank}. To limit the extent of jitter, the platform uses machine learning models to predict the performance of a function with colocation and place it on the least loaded node in a cluster~\cite{pred1, pred2}. Furthermore, the platform dynamically resizes the resources of a sandbox pool of a function based on the arrival rate in order to reuse the system's resources~\cite{lass}.

\begin{table*}
	\centering
	\footnotesize
    \caption{Workloads used in the paper (adapted from highly cited prior work~\cite{sebs, functionbench, ensure,
    faasprofiler}). The product of the service time and arrival rate is set to 0.9 (see
    Section~\ref{sec:eval_methodology}).}
	\label{tab:workload}
	\begin{tabular}{|p{0.1\linewidth}|p{0.19\linewidth}|p{0.4\linewidth}|p{0.05\linewidth}|p{0.05\linewidth}|p{0.07\linewidth}|}
		\hline
		{\bf Application category} & {\bf Workloads}                     & {\bf Description}                                                                                     & {\bf Service time (sec)} & {\bf Arrival rate (RPS)} & {\bf Latency category}  \\
		\hline
		Web                  & Markdown Renderer (MR)~\cite{faasprofiler}        & Renders a markdown page as a HTML page.                                                          & 0.125                                                          & 7.20	& LS                                                              \\
		\hline
		Web                  & Email Generator (EG)~\cite{ensure}          & Sends an auto-generated email.                                                                  & 0.220                                                          & 4.09 & LS                                                              \\
		\hline
		Web                  & Stock Analyzer (SA)~\cite{ensure}           & Analyzes stocks of a company for a given duration.                                              & 0.400                                                           & 2.25 & LS                                                              \\
		\hline
		Security             & Binary Scanner (BS)~\cite{lass}           & Scans a binary for malware traces using YARA rules.                                               & 0.160                                                          & 5.63 & LS                                                              \\
		\hline
		IoT                  & Object Detector (OD)~\cite{lass}          & Detects an object from a webcam feed using
        the SqueezeNet model.                                 & 0.300                                                          & 3.00 & LS                                                            \\
		\hline
		Multimedia           & Video Processor (VP)~\cite{sebs}          & Converts a video into grayscale.                                                                & 13.100                                                          & 0.07 & LD                                                          \\
		\hline
		Multimedia           & Image Resizer (IR)~\cite{faasprofiler, ensure}            & Resizes an image                                                                                & 0.003                                                          & 300.00 & LD                                                            \\
		\hline
		Scientific & Pagerank Checker (PC)~\cite{sebs}         & Computes the page rank of a page in a graph based on the Barabasi-Albert model.                     & 1.450                                                          & 0.62 & LD                                                           \\
		\hline
		Visualization        & DNA Visualizer (DV)~\cite{sebs}           & Visualizes DNA using the Python squiggle package.                                               & 0.450                                                          & 2.00 & LD                                                           \\
		\hline
		Machine Learning     & Product Review Analyzer (PRA)~\cite{ensure} & Generates a regression model using the logistic regression model trained on Amazon product reviews. & 8.750                                                             & 0.10& LD                                                          \\
		\hline
	\end{tabular}
\end{table*}

\begin{table}
	\centering
	\footnotesize
	\caption{System configuration}
	\label{tab:system}
	\begin{tabular}{|l|l|l|l|}
		\hline
		\multicolumn{4}{|c|}{\bf Hardware settings}                          \\
		\hline
		Processor    & \multicolumn{3}{l|}{Intel Xeon E-2186G CPU, 3.80\ GHz} \\
		\hline
		CPUs        & 1 Socket, 6 cores    & DRAM            & 32 GB          \\
		\hline
		\multicolumn{4}{|c|}{\bf Software settings}                                 \\
		\hline
		Linux Kernel & 5.14                 &    ASLR         & Off             \\
		\hline
	\end{tabular}
\end{table}

\subsection{Reinforcement Learning}
Reinforcement learning (RL) is a machine learning technique that employs an intelligent agent that takes actions in an uncertain environment, where the aim is to maximize the cumulative reward: the sum of the incentives received by the agent. The system is modeled as a Markov decision process (MDP)~\cite{rl} comprising a set of states representing the environment ($\mathcal{S}$), a set of actions that can be performed by the agent ($\mathcal{P}$), a function that denotes the probability of transitioning from state $s$ to state $s'$ for a given action $a$, and a reward function that provides the reward for the aforementioned state transition. The objective of any RL scheme is to come up with a policy $\pi$ that determines the probability distribution of actions in a given state such that the expected value of the cumulative reward is maximized.

The designer typically provides a description of states, a list of actions, and a reward function. The function $\pi$ is computed automatically by the RL library; it basically uses a search technique. A common technique is the policy iteration technique~\cite{rl} that relies on Monte Carlo simulation. Sadly, this technique is not suitable for a scenario like ours where the variance in the rewards is high~\cite{rl}. This is because we see a lot of contention and non-determinism in our system. A newer set of approaches employs temporal difference methods~\cite{td_1, td_2} that additionally associate a {\em value} with each state. It is in the line of classical fixed-point approaches where the state-value function represents the expected cumulative reward that an agent will accumulate by starting from a given state. In this set, \methodname employs the actor-critic scheme because it has good convergence properties~\cite{acrl}. Again, within this subset, the advantage actor-critic (~\emph{A2C-RL}) method was found to be the best because of its superior results~\cite{a2crl}.

\subsubsection{A2C-RL}
\label{sec:a2crl}
\begin{figure}[!ht]
	\centering
	\includegraphics[width=0.6\linewidth]{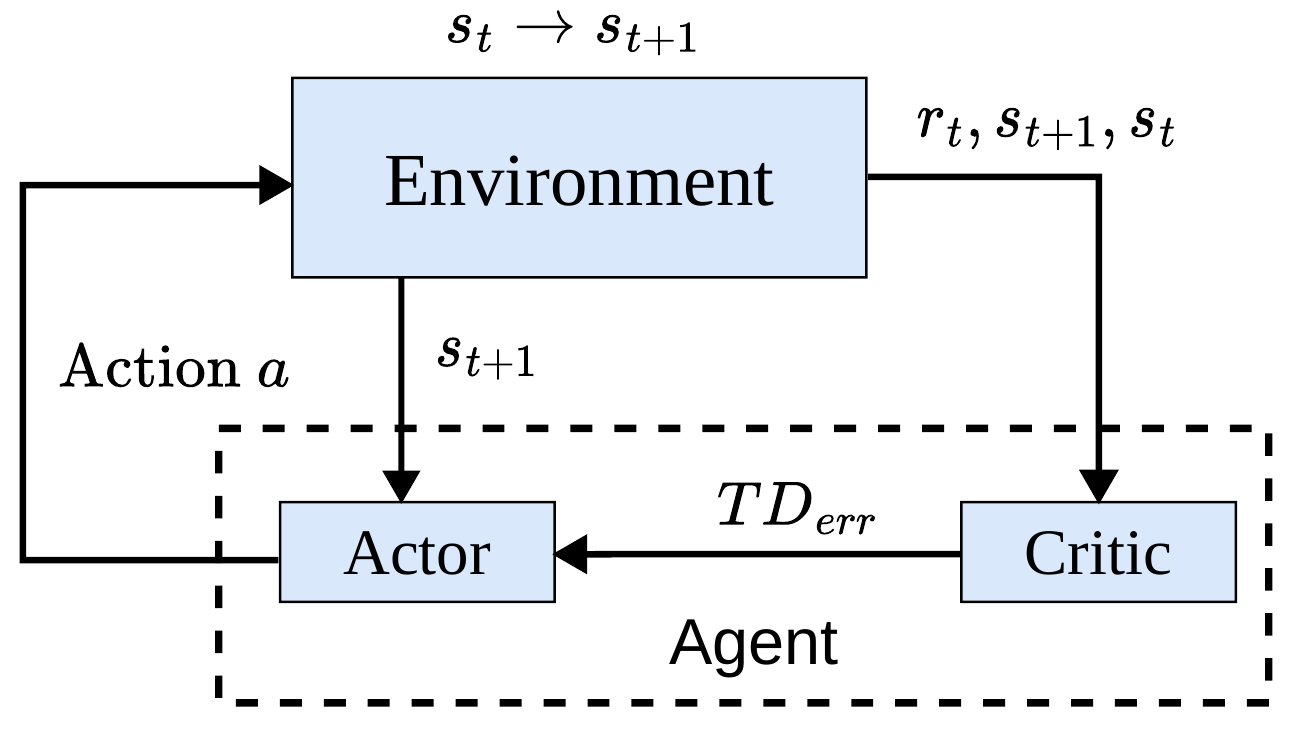}
	\caption{Design of an actor-critic reinforcement learning scheme}
	\label{back:rl_design}
\end{figure}

The \emph{A2C-RL} comprises the following components: the environment, the critic, the actor and the agent (refer to Figure~\ref{back:rl_design})~\cite{reinforcement}. When the current state of the environment is $\mathbf{s_t}$, the agent applies an action $a$ to the environment. Consequently, the state of the environment changes to $\mathbf{s_{t+1}}$. While using the model, also known as {\em exploitation}, we need to optimally find $s_{t+1}$. A2C-RL methods derive their correctness from optimal control policies based on the classical Bellman equation~\cite{bellman}, where we just greedily move to that state which has the highest state-value function. A locally optimal action is also the most globally optimal.

This basically means that in the training phase (also known as the {\em exploration phase}), we need to get precise estimates of the state-value function. The state-value function for any state is initialized to a random value. On every training step, we update $V(s)$ as follows: $V(s) += \alpha \times TD_{err}$. Here, $V(s)$ is the state-value function for state $s$, $\alpha$ is the learning rate, and $TD_{err}$ is the temporal difference error term, which is given by Equation~\ref{equ:td_error}. 

\begin{equation}
	\label{equ:td_error}
        TD_{err} =  r_{t} + \gamma\mathcal{V}(\mathbf{s_{t+1}}) - \mathcal{V}(\mathbf{s_{t}}) 
\end{equation}
Here, $r_t$ is the instantaneous reward and $\gamma$ ($\in [0,1)$) is an ageing factor. RL algorithms need not have an offline exploration phase. In our case, they are trained dynamically where we interleave periods of exploration and exploitation. $\alpha$
is set to 0.0001, and $\gamma$ is set to 0.99 (on the lines of~\cite{rl_lr1, rl_lr2}).

\section{Characterization of a Serverless Framework}
\label{sec:characterization}
In this section, we use LaSS~\cite{lass}, a state-of-the-art serverless framework, to analyze the jitter in the execution latency when we run one LS application alongside an LD application. It regulates the resources of the hosted applications based on the arrival rate and service time of the applications.

\subsection{Evaluation Methodology}
\label{sec:eval_methodology}

We evaluate LaSS~\cite{lass} on a single server and study the CPU contention and the microarchitectural-level interference suffered by the container (sandbox) processes of an application. For ease of analysis, we use a pair of an LS application and an LD application. The workloads comprise popular real-world serverless applications, which are summarized in Table~\ref{tab:workload}. The system configuration is shown in Table~\ref{tab:system}. We assume that the request arrival rate follows the Poisson distribution and the service time is exponentially distributed (similar to LaSS~\cite{lass}). We model the system using the $M/M/c/FCFS$ queuing system with $c$ servers to process incoming requests. For a serverless function with a mean service time ($\mu$) and a mean request arrival rate ($\lambda$), the queue utilization ($\rho$) for a duration of $\mathcal{T}$ seconds can be formalized as the fraction of the total time servers are being utilized by incoming requests (see Equation~\ref{equ:utilization}). Prior work~\cite{queue1, queue2, queue3} in the domains of fog computing, cloud computing and IoT has shown that real-world applications exhibit a queue utilization that is in the ballpark of 0.9. Henceforth, the request drop rate increases significantly. Therefore, for each application, the request arrival rate ($\lambda$) is chosen such that the queue utilization ($\rho$) is $0.9$.

\begin{equation}
	\label{equ:utilization}
 \rho = \frac{\text{Total service time}}{\mathcal{T}} = \frac{\mathcal{T} \times \lambda \times \mu}{c\mathcal{T}}  
\end{equation}

To measure the jitter in the latency, we employ the following statistical measures: interquartile range (IQR), variance, and mean (similar to \cite{iqr1,iqr2}). The IQR value is calculated as the difference between the third quartile $Q_3$ and the first quartile $Q_1$ of a distribution. The IQR value eliminates the top 25\% and the bottom 25\% of the distribution, hence it is not very sensitive to outliers. {\bf Note that} we implemented all the standard measures to reduce jitter (detailed by Pradipta et al.~\cite{jitter}). In our study, we compare the value of the aforementioned metrics with respect to when an application is executed in isolation.

\begin{figure}[!htb]
	\centering
	\begin{subfigure}[b]{0.525\linewidth}
		\captionsetup{justification=centering}
		\includegraphics[width=\linewidth]{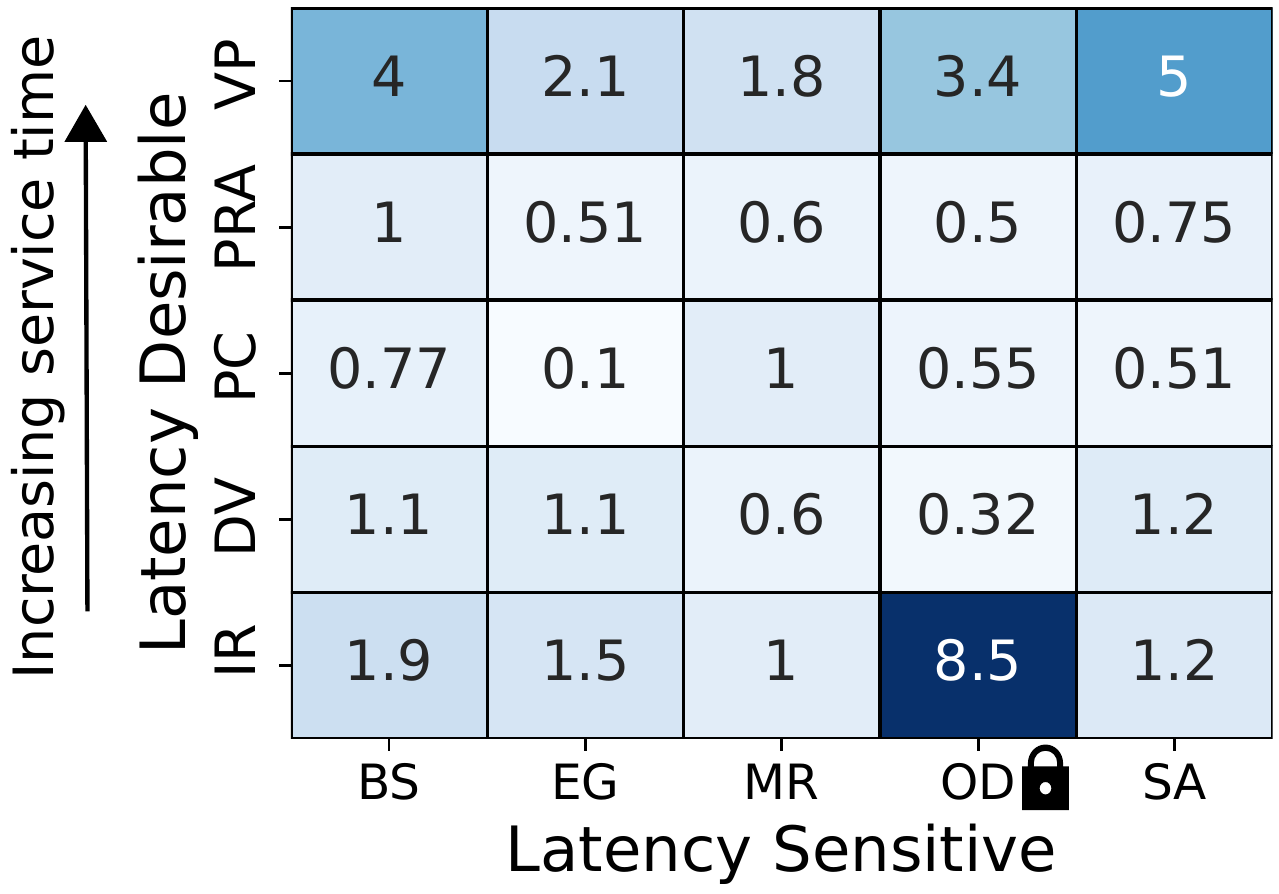}
		\caption{ Normalized IQR of the execution latency}
		\label{motiv:iqr}
	\end{subfigure}
	\hfil
	\begin{subfigure}[b]{0.405\linewidth}
		\captionsetup{justification=centering}
		\includegraphics[width=\linewidth]{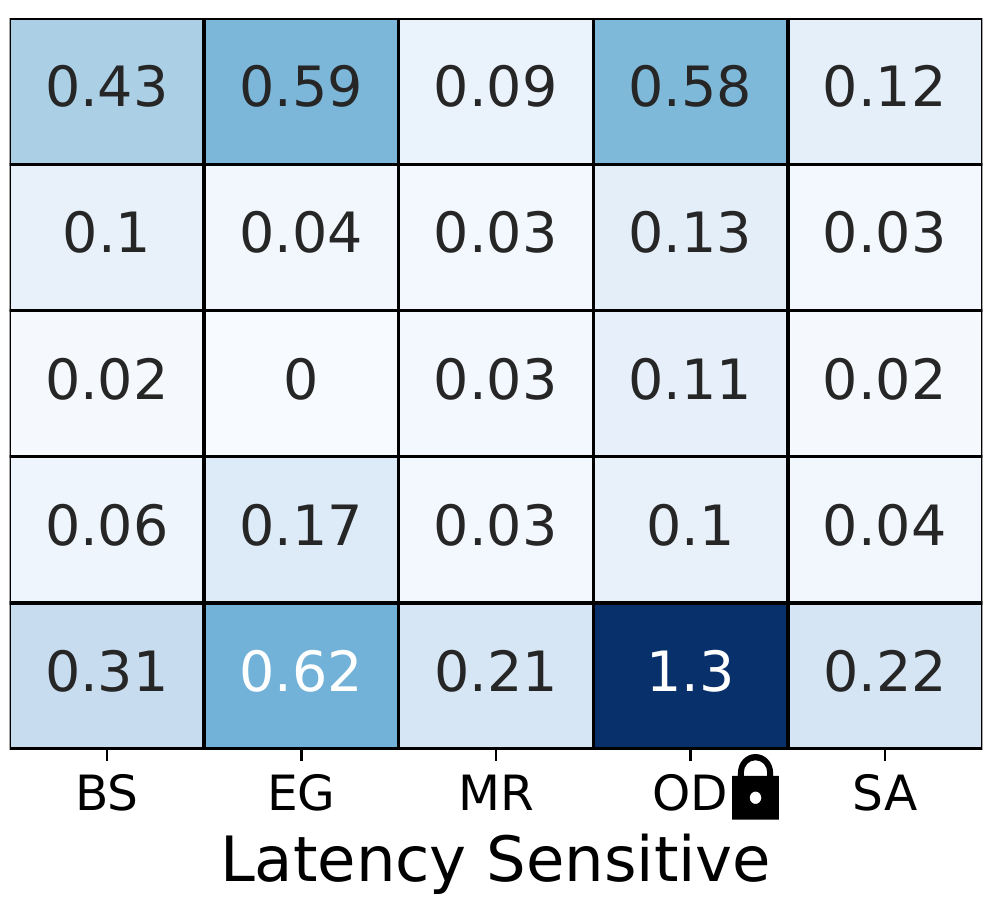}
		\caption{Normalized mean execution latency }
		\label{motiv:mean}
	\end{subfigure}
	\caption{Degradation in the IQR value and the mean of the execution latency of an LS application when it is colocated with an LD application (normalized to an isolated execution of the LS application).}
	\label{motiv:exec_variation}
\end{figure}

\begin{figure}[!htb]
	\centering
	\includegraphics[width=\columnwidth,keepaspectratio]{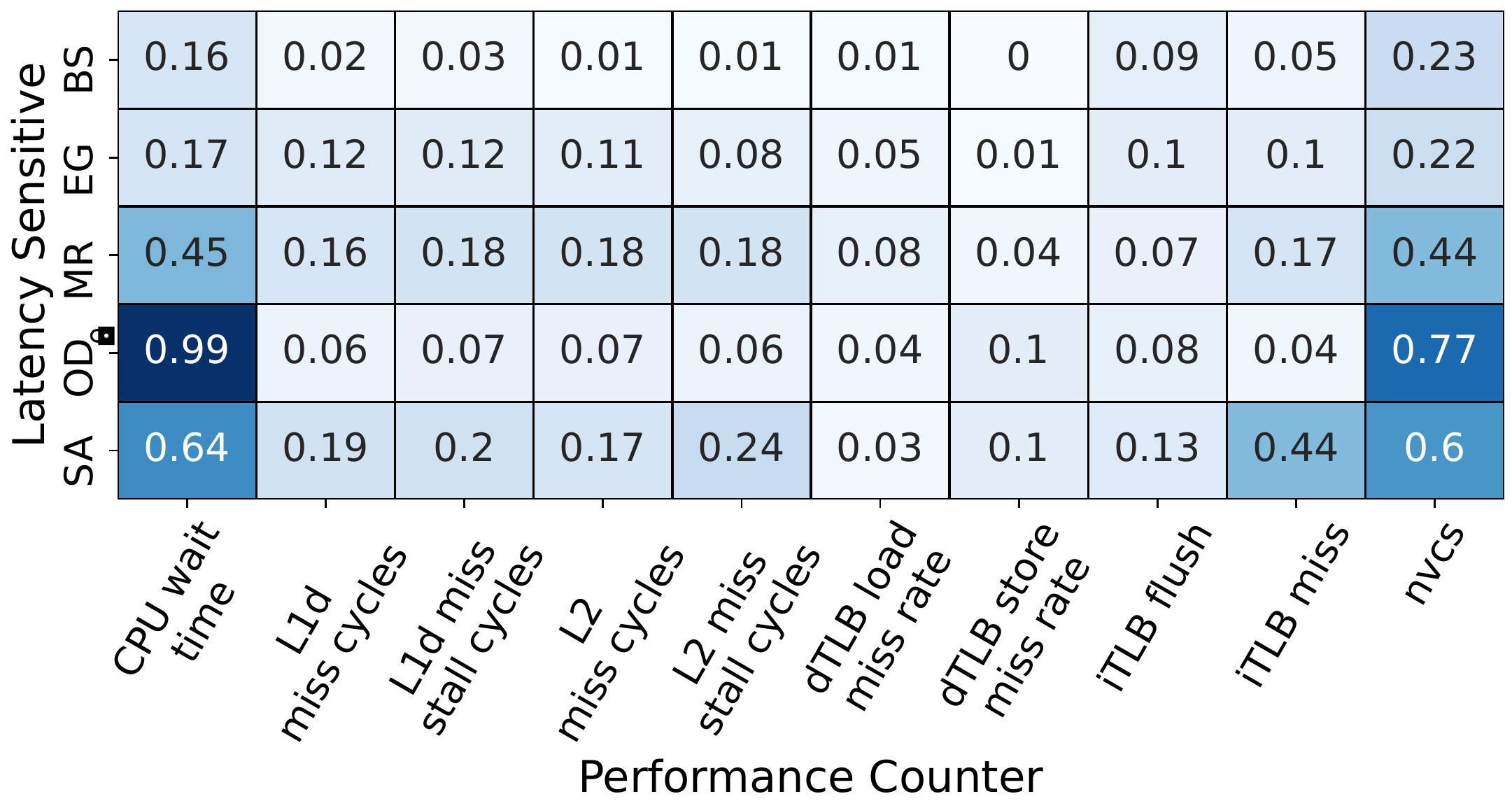}
	\caption{Correlation of HPEs and OS events with the execution latency of the LS application}
	\label{motiv:correlation}
\end{figure}

\begin{figure*}[!htb]
	\centering
	\begin{subfigure}[b]{0.285\linewidth}
		\captionsetup{justification=centering}
		\includegraphics[width=\linewidth]{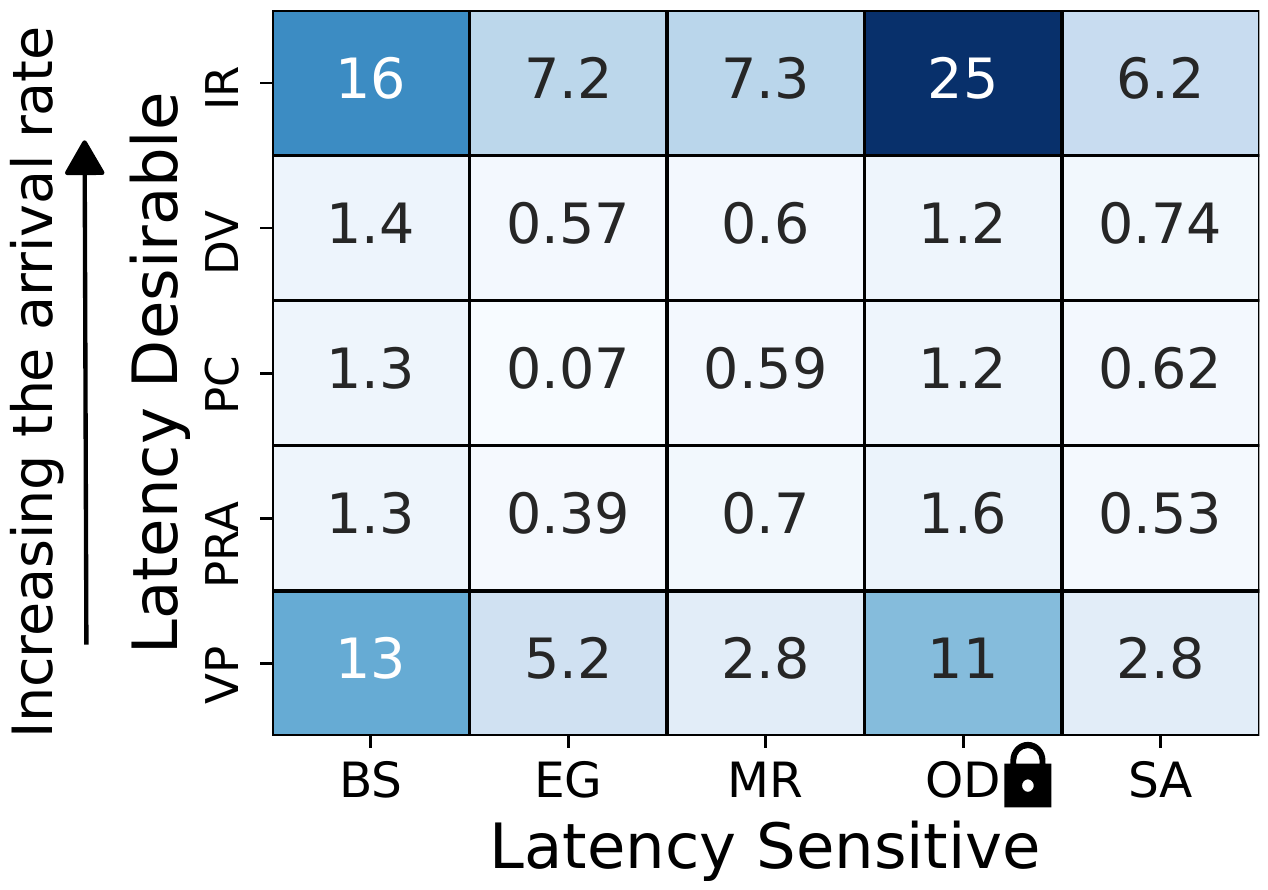}
		\caption{CPU wait time}
		\label{motiv:wait_time}
	\end{subfigure}
	\hfil
	\begin{subfigure}[b]{0.225\linewidth}
		\captionsetup{justification=centering}
		\includegraphics[width=\linewidth]{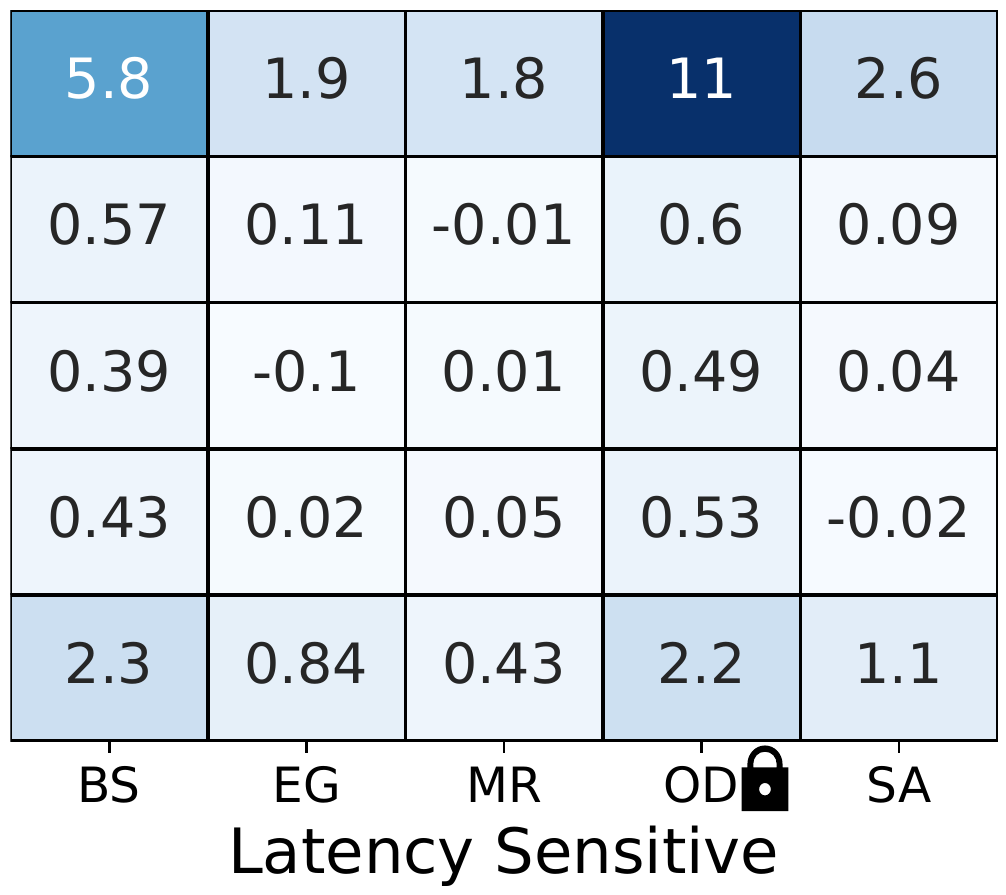}
		\caption{\emph{nvcs} events}
		\label{motiv:nvcs_events}
	\end{subfigure}
	\hfil
	\begin{subfigure}[b]{0.225\linewidth}
		\captionsetup{justification=centering}
		\includegraphics[width=\linewidth]{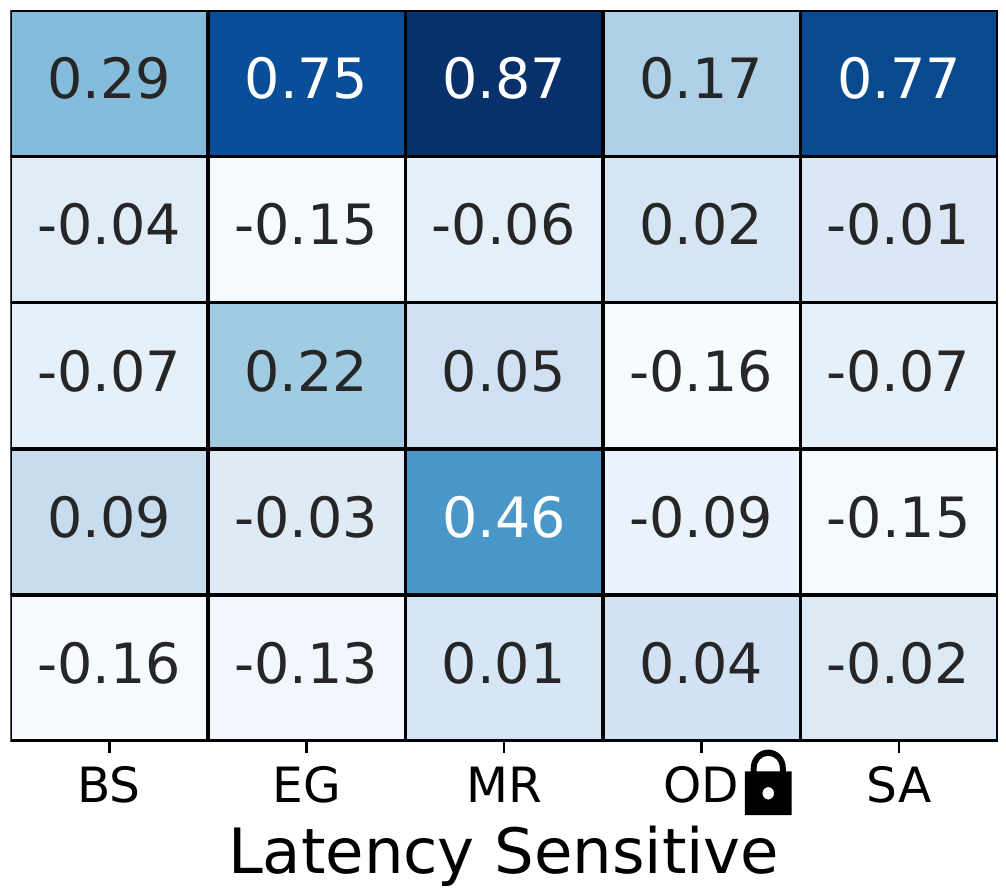}
		\caption{iTLB flushes}
		\label{motiv:itlb_flush}
	\end{subfigure}
	\hfil
	\begin{subfigure}[b]{0.229\linewidth}
		\captionsetup{justification=centering}
		\includegraphics[width=\linewidth]{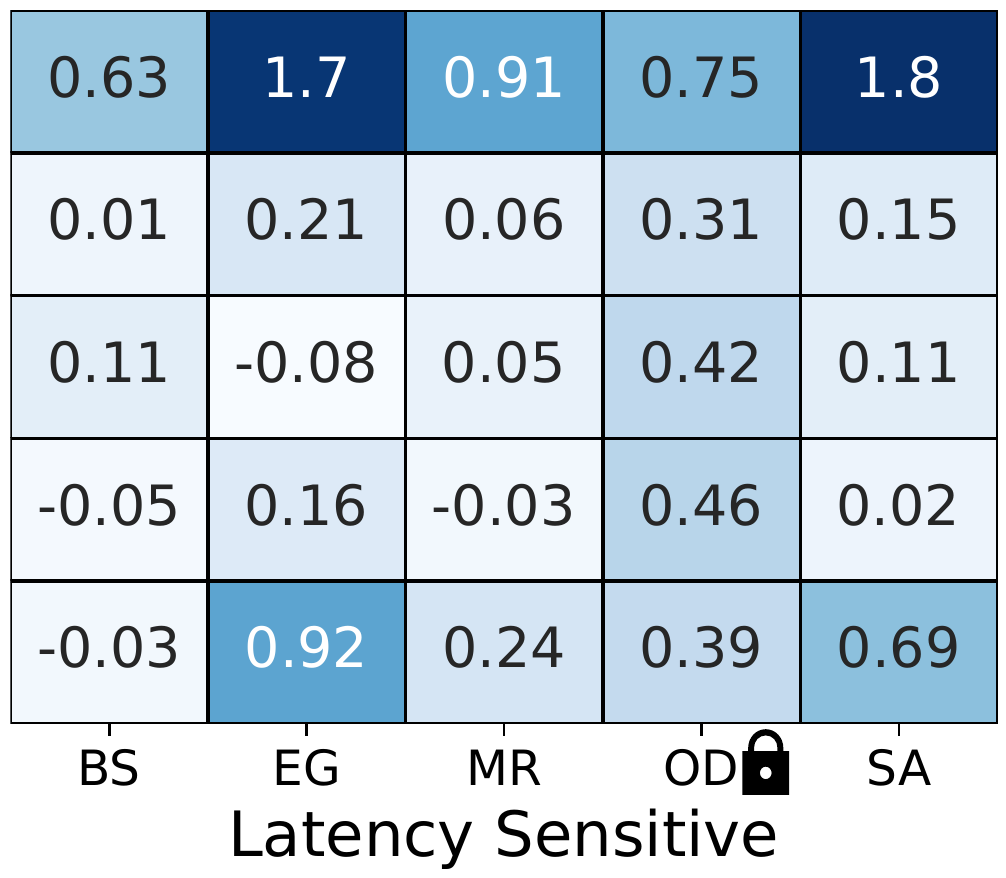}
		\caption{iTLB misses}
		\label{motiv:itlb_miss}
	\end{subfigure}
	
	\caption{Normalized performance counter values and execution statistics of the LS application (with respect to when the LS application is executing in isolation)}
	\label{motiv:cpu_variation}
	
\end{figure*}

\subsection{Quantifying the Interference}
\label{motiv:variability}
To capture the effect of CPU contention and the microarchitectural-level interference suffered by an LS application, we measure the IQR value and the mean of the execution latency (time spent by a request executing inside a container). We execute each of the LS applications along with each of the LD applications. In Figure~\ref{motiv:exec_variation}, we show that the IQR value and the mean of the execution latency of an LS application increase by up to 8.5$\times$ and 1.3$\times$, respectively, with colocation compared to itself when executed in isolation. Note that each of the subfigures contains ratios. The lock sign beside the OD application indicates that it uses locks.

When an LS application (except object detector (OD)) is colocated with the video processor (VP) application, we observe the maximum degradation in the IQR value of the execution latency compared to other LD applications. This is because the service time (= $12$ seconds) of the VP application is higher than that of other LD applications. Furthermore, we observe an increase of at least $10\%$ in the mean execution latency of the object detector (OD) application (as shown in Figure~\ref{motiv:mean}), regardless of the colocated application, due to the file-level futex lock on \emph{libcaffe2.so} of the python-torch library that is shared across the container processes.

\begin{tcolorbox}[top=4pt,left=0pt,right=4pt,bottom=0pt]
	\begin{enumerate}
        \item With colocation, an LS application can suffer an increase in the IQR value and the mean of the execution latency by up to 8.5$\times$ and 1.3$\times$, respectively.
 		\item If an LS application employs a locking mechanism, then the mean execution latency can increase by at least $10\%$ with colocation.
	\end{enumerate}
\end{tcolorbox}

\subsection{Understanding the Interference}
We collect hardware performance events (HPEs) provided by Intel Xeon processors and CPU-usage events from the \emph{proc} file system during the execution of an application's request. The CPU-usage events capture the per-process CPU wait time (time spent by a process and its threads waiting in the run queue of a core) and the total number of non-voluntary context switch events (\emph{\#nvcs}). Subsequently, we perform statistical analyses to select a set of HPEs and CPU-usage events that are highly correlated with the execution time of an application using Pearson correlation coefficients. 

In Figure~\ref{motiv:correlation}, we show that the CPU wait time and the \emph{\#nvcs} events are highly correlated with the execution latency of an LS application's request. Furthermore, we show that the effect of HPEs related to dTLB and cache accesses is minimal. For the object detector (OD) application, the CPU wait time is strongly correlated with the execution latency as compared to the \emph{\#nvcs} events. This is because container processes are waiting to acquire the futex lock on the shared file (\emph{libcaffe2.so} of the python-torch library). In the case of the stock analyzer (SA) application, we observe that the total number of iTLB miss events ({\em \#iTLB\_misses}) is strongly correlated to the execution latency of a request.

\subsubsection{Analyzing the CPU Contention}
\label{sec:cpu_wait}

In Figures~\ref{motiv:wait_time} and \ref{motiv:nvcs_events}, we show that when an LS application is colocated with an LD application, the CPU wait time and the \emph{\#nvcs} events increase by up to 25$\times$ and 11$\times$, respectively. This is because the operating system treats container processes of an application as regular processes and uses the regular SCHED\_OTHER scheduling policy. As a result, LS applications experience jitter in the execution latency.

When an LS application is colocated with the image resizer (IR) application, we observe an increase  in the CPU wait time and the \emph{\#nvcs} events by $6.2\times$ to $25\times$ and by $1.8\times$ to $11\times$, respectively. This is because the request arrival rate of the IR application is higher than that of other LD applications, at 300 requests per second (RPS). Hence, the container processes of the IR application contend quite heavily for CPU resources. We observe a higher degree of CPU contention in OD because of the same reasons (file-level lock on libcaffe2.so).

\begin{tcolorbox}[top=4pt,left=0pt,right=4pt,bottom=0pt]
\begin{enumerate}
	\item The CPU wait time and the \emph{\#nvcs} events suffered by an LS application are dependent on the request arrival rate of an LD application (E.g.: IR).
	\item If an LS application such as OD employs a locking mechanism, then the jitter in the execution latency can be partially accounted for by the application's container processes  waiting to acquire the lock.
\end{enumerate}
\end{tcolorbox}

\subsubsection{Analyzing iTLB Behavior}
\label{sec:itlb}
If a serverless application has poor code locality, then it will suffer from memory stalls due to a large number of iTLB misses~\cite{code_locality}. In this subsection, we discuss the impact of iTLB behavior on the execution latency by profiling the {\em \#iTLB\_flushes} and {\em \#iTLB\_misses}. In Figures~\ref{motiv:itlb_flush} and \ref{motiv:itlb_miss}, we show that when an LS application is colocated with an LD application, the {\em \#iTLB\_flushes} and  {\em \#iTLB\_misses} increase by up to $87\%$ and $180\%$, respectively. As a result, contention in the iTLB causes jitter in the following applications: stock analyzer (SA), email generator (EG), and markdown renderer (MR).

\begin{tcolorbox}[top=4pt,left=0pt,right=4pt,bottom=0pt]
In some LS applications, lack of code locality manifesting in poor iTLB behavior contributes to performance degradation; we see an increase in {\em \#iTLB\_flushes} and {\em \#iTLB\_misses}  by up to 87$\%$ and 180$\%$, respectively. These are a direct result of destructive interference due to colocation.
\end{tcolorbox}

\subsection{Effect of the Input Size}
\label{sec:cov_ipc}

\begin{figure}[!htb]
	\centering
	\includegraphics[width=0.9\linewidth,keepaspectratio]{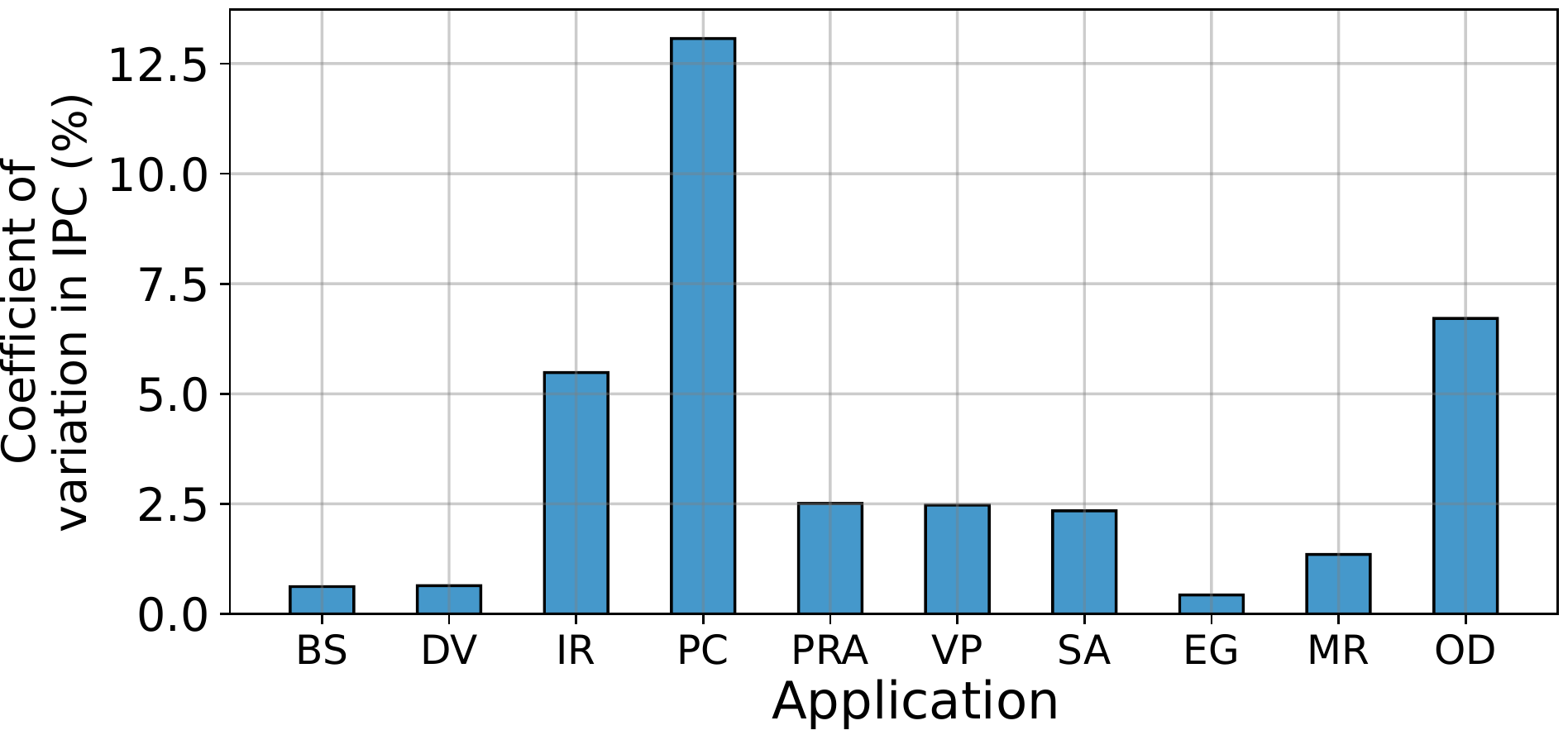}
	\caption{The $\sigma/\mu$ values (coefficient of variation) of the IPC of applications as the input size changes}
	\label{motiv:cov_ipc}
\end{figure}

In a serverless computing platform, the input is provided by an external user. To measure the slowdown in an application (w.r.t. native execution), we should employ a metric that varies minimally with the input size of the application. We measure the instructions-per-cycle (IPC) metric of an application by varying the size of the input. To do this experiment, we considered larger inputs (up to $5 \times$ more). For a lack of space, we are not listing all the sources of the new inputs, however, the same can be provided on request. Just as an example, for the image resizer (IR), we just downloaded random images from Google Images of the appropriate size, or for the Video Processor (VP)  we downloaded a random video from YouTube. The choice did not make a difference. As we can see in Figure~\ref{motiv:cov_ipc}, the mean IPC across the inputs remained stable ($\sigma/\mu$ within 3.6\% on an average).

\section{Motivation}
\label{sec:motivation}

In this section, we motivate the need for a framework that intelligently regulates the allocated CPU resources to minimize jitter in the latency of latency-sensitive (LS) applications.

\begin{figure}[!htb]
	\centering
		\captionsetup{justification=centering}
		\includegraphics[width=0.9\columnwidth]{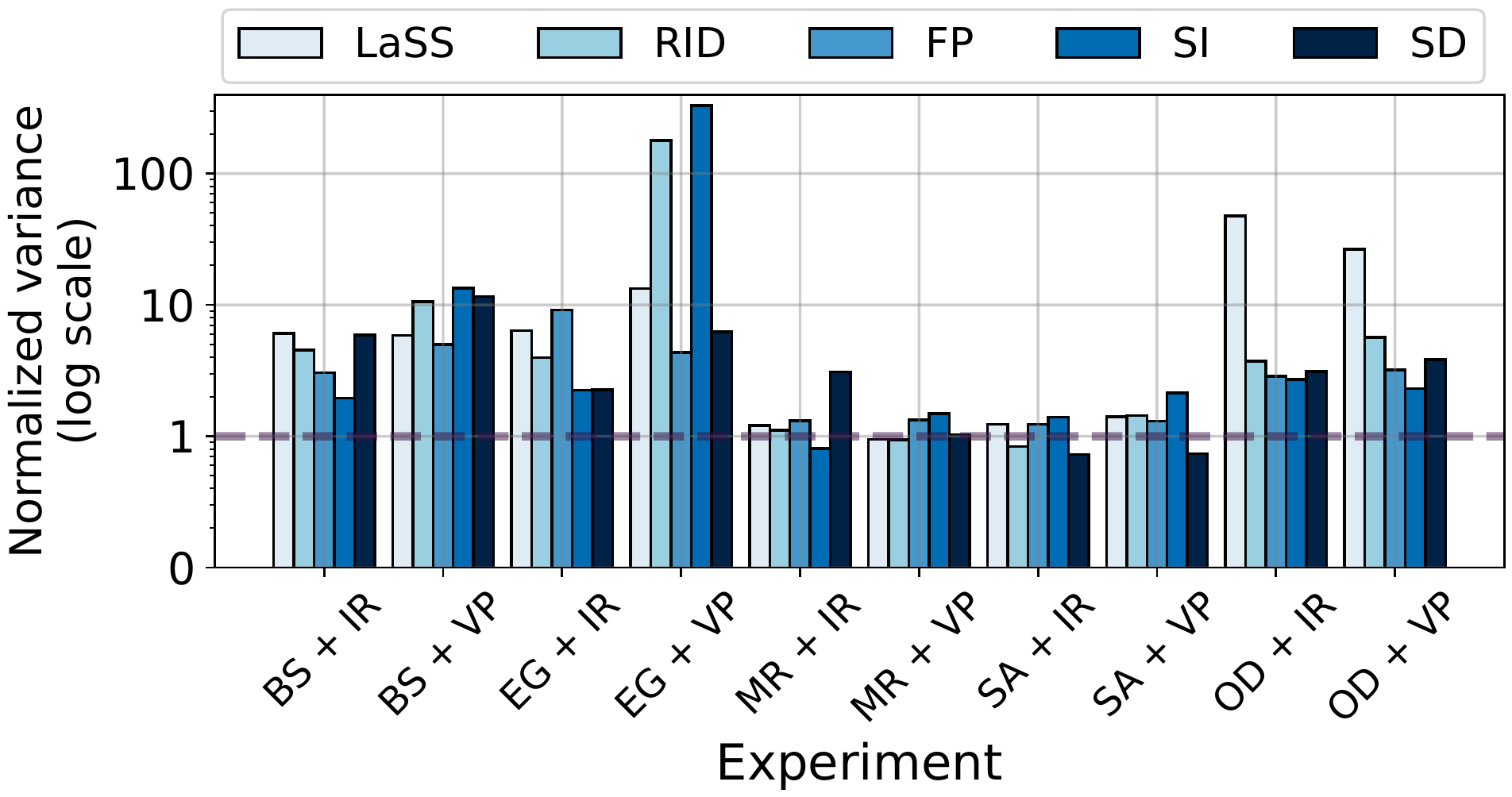}
		\caption{Normalized variance in the response latency of LS applications}
	\label{motiv:prio_regulation}
\end{figure}

\subsection{Default Knob: Assigning Priority}
To reduce CPU contention, the scheduling policy of the container processes executing an LS application can be changed from SCHED\_OTHER to SCHED\_RR (round robin real-time scheduling policy). We found SCHED\_FIFO to be quite sub-optimal. Now, the OS can
prioritize the container processes of an LS application over other applications running with the  SCHED\_OTHER policy (the time slice is 100 ms~\cite{schedrr}). Recall that Linux real-time priorities can vary from 1 to 99 (highest priority).

Let us discuss the impact of the SCHED\_RR scheduling policy on an LS application when colocated with an LD application. For this study, we use the following priority assignment schemes for the container processes of an LS application: \circled{1} randomly increasing or decreasing priority (\emph{RID}), \circled{2} fixed priority of 80~\cite{jitter} (\emph{FP}), \circled{3} strictly increasing priority (\emph{SI}), and \circled{4} strictly decreasing priority (\emph{SD}). {\bf Note that for scheduling priorities, higher is not always better~\cite{jitter}.}

In Figure~\ref{motiv:prio_regulation} we plot the variance in the response latency for LaSS and four of the priority assignment schemes. They are all normalized to the isolated execution case. IR and VP were chosen as LD applications because of their high arrival rate and service time, respectively. In the figure, OD+IR and OD+VP are outliers mainly because of the futex lock issue. For the rest, the variance increases by roughly 1.5-6$\times$ across schemes. Large values are not uncommon though: more than 10$\times$ in EG+VP. Among the priority schemes, no clear winner is emerging and we are seldom close to the isolated case. 

\begin{tcolorbox}[top=4pt,left=0pt,right=4pt,bottom=0pt]
    The SCHED\_RR scheduling policy allocates more CPU time to LS applications, but setting the priority assigning scheme for the container processes to minimize the jitter is not trivial. 
\end{tcolorbox}

Note that the FP scheme (fixed at 80) is not always the best. The reason for this is that in modern serverless applications, there is a very complex interaction between the container processes, the framework, and the OS. As a result, increasing the priority of some processes hurts the application in the long run because kernel threads and daemons do not run that frequently (also observed in~\cite{highprio1, highprio2}).

\subsection{Setting the Physical CPU Affinity}
\label{sec:motiv_lock}
To minimize the contention in the iTLB, we can restrict the container processes of an LS application to a set of physical cores. This will improve iTLB and i-cache performance. Using the same setup as before, let us partition the available physical CPU cores in an $m:n$ ratio, where $m$ represents the number of physical cores allocated to the LS application and $n$ represents the number of physical cores allocated to the colocated LD application.

\begin{figure}[!htb]
	\centering
	\captionsetup{justification=centering}
	\includegraphics[width=\columnwidth]{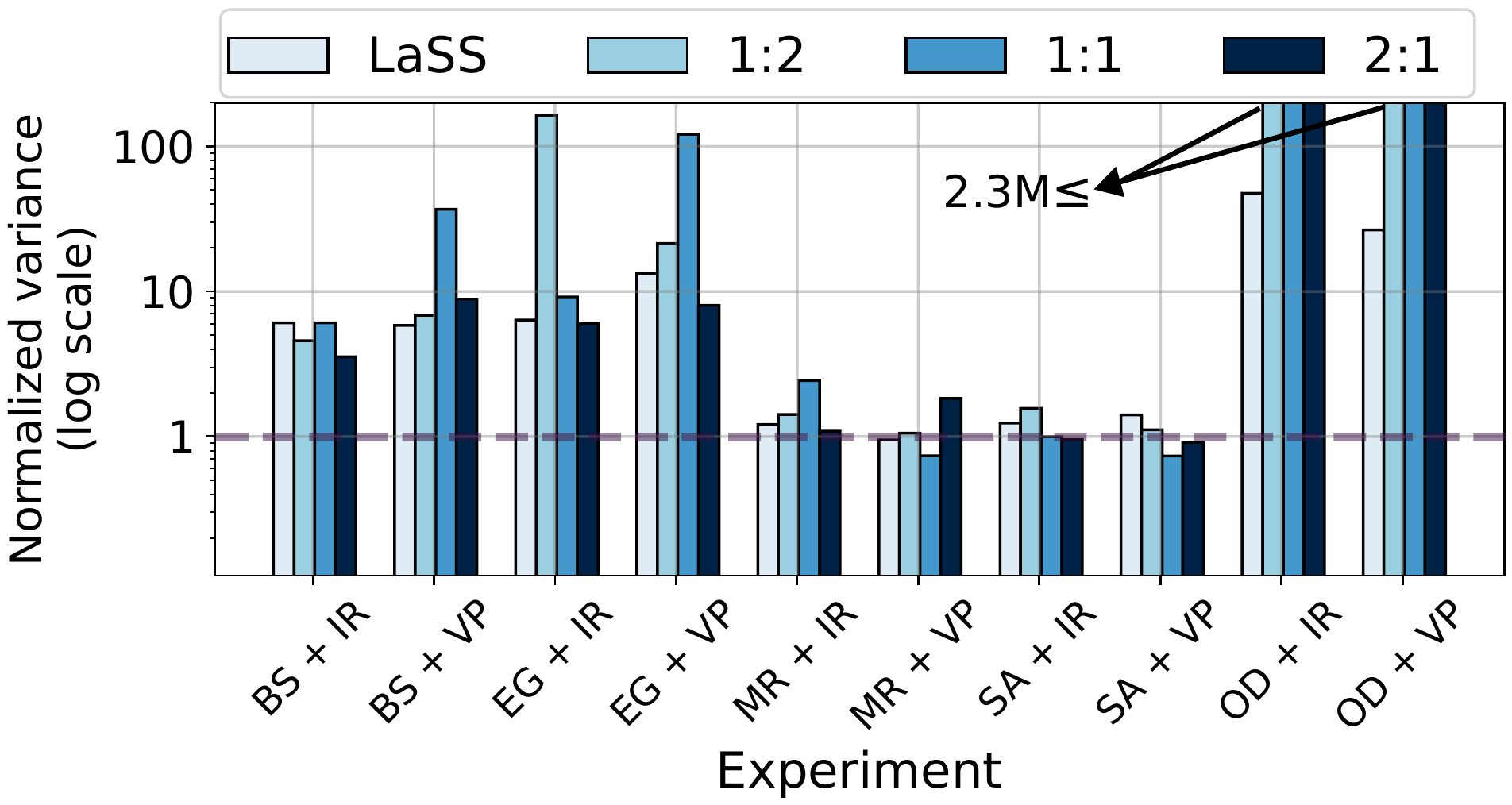}
	\caption{Normalized variance in the response latency of LS applications. Varying ratios of $\langle$\#LS
		cores$\rangle$:$\langle$\#LD cores$\rangle$ }
	\label{motiv:affin_regulation}
\end{figure}

In Figure~\ref{motiv:affin_regulation}, we show that allocating dedicated physical cores improves the variance in the response latency of LS applications (except for the OD+X applications) by an average $17.88\%$ compared to LaSS~\cite{lass}. This does not convey the full picture. We have good results for 4 out of the remaining 8 combinations and inferior results for the rest. For the OD+IR and OD+VP combinations, the degradation is quite large when the physical CPU affinity is set, compared to LaSS~\cite{lass}. As a result, a more intelligent mechanism is required to set the physical CPU core allocation of an LS  application.

\begin{tcolorbox}[top=4pt,left=0pt,right=4pt,bottom=0pt]
Setting core affinities does not help all the time, regardless of the $m:n$ ratio. It is particularly difficult for applications like OD that access futex locks. A more intelligent scheme is needed.
\end{tcolorbox}

\section{Design}
\label{sec:design}
In the previous section, we showed that simple priority setting and physical CPU core affinity setting schemes do not provide good results. Their performance varies across workload combinations. Since the primary task is decision making in an uncertain environment, reinforcement learning (RL) based techniques are naturally germane to such scenarios (similar to prior work that targets conceptually similar problems~\cite{rl_cpu1, rl_cpu2}).

\methodname uses the Advantage Actor-Critic reinforcement learning (\emph{A2C-RL}) methodology~\cite{reinforcement} to decide the \emph{scheduling policy} also referred to as \emph{Sched policy} based on the CPU contention and microarchitectural interference suffered by an LS application (as discussed in Section~\ref{sec:background}). The Sched policy is a 2-tuple of the priority of all the containers (same priority for all) and the number of dedicated CPU cores assigned to the application's container processes.

\subsection{High-Level Overview}

\begin{figure}
	\centering
	\includegraphics[width=0.8\linewidth]{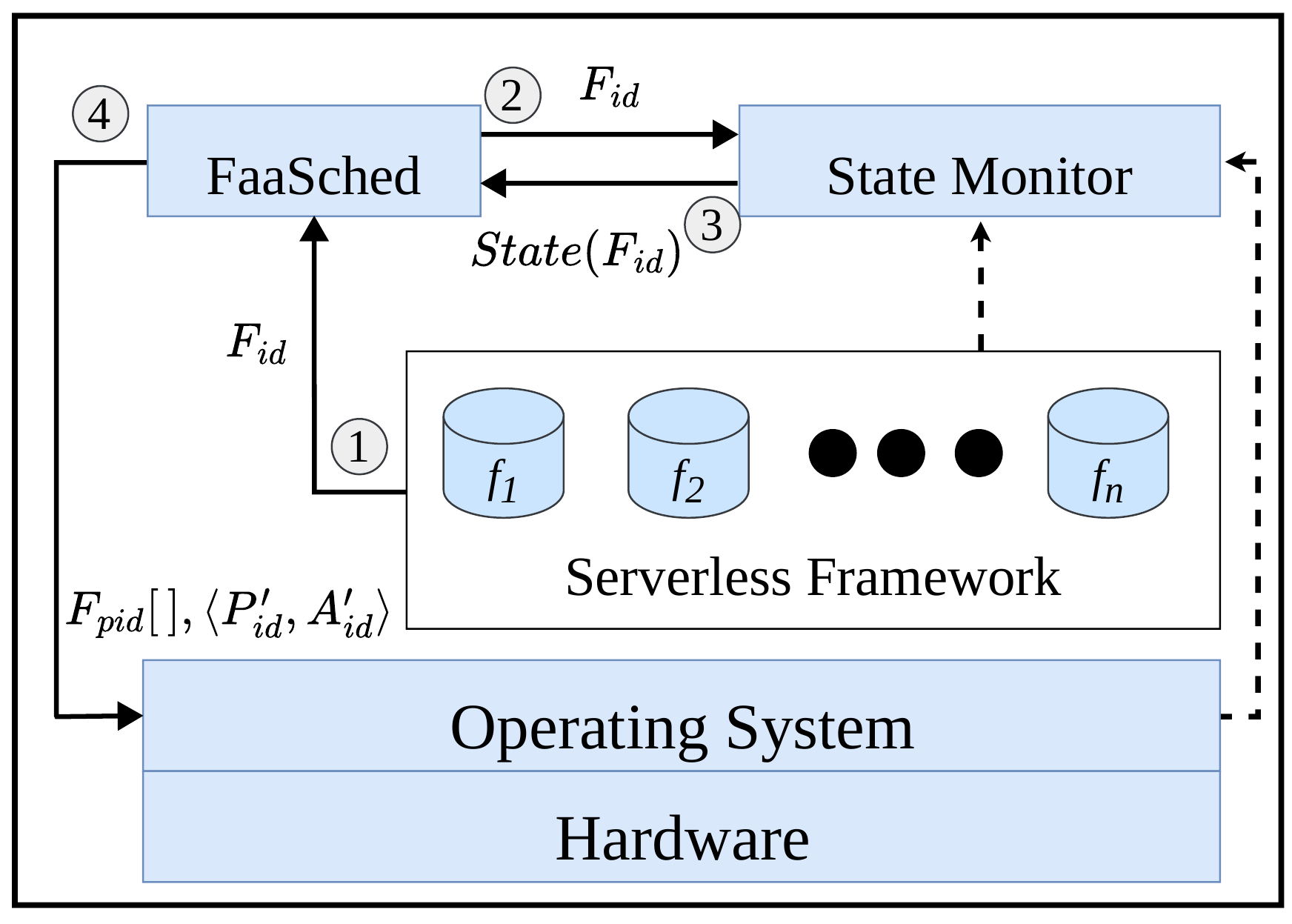}
	\caption{The high-level design of \methodname.}
	\label{design:architecture}
\end{figure}

In Figure~\ref{design:architecture}, we show the high-level design of \methodname. The serverless framework sends a request to \methodname in order to set the \emph{Sched policy} of a serverless application ($F_{id}$) (indicated as \circled{1} in Figure~\ref{design:architecture}). Subsequently, \methodname fetches the \emph{state} of the application and the entire system $State(F_{id})$ from the state monitor daemon (\circled{2} and \circled{3}). The fields of the state are shown in Table~\ref{tab:conventions}. Note that defining the state is a very tricky process in any RL scheme. We have opted for a partially observable strategy, where the state from the point of view of an application comprises some of its execution parameters/attributes and an estimate of the behavior of the rest of the system. The fields $F_{pid}[\,]$, $P_{id}$, $A_{id}$, $F_{lock}$, and $S_{cont}[\,]$ are application specific. The other terms represent the approximate behavior of the rest of the system. They will be elaborated in the subsequent sections. Note that we use the $[\,]$ symbol to indicate vectors.

We use the \emph{A2C-RL} methodology to compute a new priority $P'_{id}$ and a new physical CPU allocation $A'_{id}$ for the application -- an application encompasses all its container processes (\circled{4}). To set the priority of the container processes and the core affinities of an LS application, \methodname utilizes the \emph{chrt} and \emph{taskset} utilities in Linux~\cite{chrt,taskset}, respectively. 

\begin{table}
	\centering
	\begin{center}
		\footnotesize
		\caption{State attributes of a serverless application ($F_{id}$).}
		\label{tab:conventions}
		\begin{tabular}{|l|p{7cm}|}
			\hline
			\multicolumn{2}{|c|}{\textbf{Notations}} \\ \hline
			$F_{pid} [\,]$ & An array of process ids of the container processes (max: 7) \\ \hline
			$P_{id}$ & The single priority allocated to the container processes  \\ \hline
			$A_{id}$ & The single physical CPU core allocation of the container processes. Represents the no. of CPU cores allocated.  \\ \hline
			$F_{lock}$ & A Boolean variable denoting whether an application uses futex locks. \\ \hline
			$S_{cont}[\,]$ & A vector representing the CPU wait time (in secs), \emph{\#nvcs} events and and \#iTLB\_misses. \\ \hline
			$S_{fair}$ & A float value denoting the slowdown in the serverless functions (w.r.t. when executed in isolation) \\ \hline
			$P_{low}$ & The number of container processes that use the SCHED\_RR scheduling policy and their priority is $\le$ $P_{id}$ \\ \hline
			$P_{high}$ & The number of container processes that use the SCHED\_RR scheduling policy and their priority is $> P_{id}$ \\ \hline
			$A_{other}$ & The total number of cores currently allocated to other LS applications running on the system  \\ \hline
		\end{tabular}%
	\end{center}
\end{table}

\subsection{State Monitor}
\label{sec:monitor}
The state monitor daemon is responsible for collecting the \emph{state} of an  application+system and providing it to \methodname (\circled{2} and \circled{3}). It collects CPU-usage events (the CPU wait time and \emph{nvcs} events) and the total number of iTLB miss events suffered by the container processes of an LS application. To detect if LS applications monopolize CPU resources, it monitors the degradation in instructions per cycle (IPC) of container processes associated with applications (with respect to when executed in isolation) (discussed in Section~\ref{sec:monitor_fair}). The stability of the IPC metric of a serverless application with a change in the input size was discussed in Section~\ref{sec:characterization}. Hence, the degradation is well defined.

\subsubsection{Contention}
\label{sec:monitor_cont}
In Section~\ref{sec:characterization}, we showed that CPU-usage events and iTLB miss events of an LS application are the primary sources of jitter. The state monitor daemon represents these events ($\langle CPU\_wait\_time, \#nvcs, \#iTLB\_misses \rangle$) as a 3-D vector; this vector {\bf represents the contention} and is referred to as $S_{cont}[\,]$.

\subsubsection{Fairness}
\label{sec:monitor_fair}
For each application, we find its slowdown by calculating the ratio of its IPC when running with colocation ($IPC^{shared}$) to its IPC when running in isolation ($IPC^{alone}$). The latter is supplied by the application developer; refer to Section~\ref{sec:cov_ipc}. Subsequently, we define the fairness metric $S_{fair}$ as the ratio of the minimum slowdown to the maximum slowdown across all the container processes of applications that are colocated in the system (ideally it should be one) (similar to \cite{xchange}).

\begin{equation}
	\label{equ:fairness}
	S_{fair} = \frac{min_{i} \frac{IPC_{i}^{shared}}{IPC_{i}^{alone}}}{max_{j} \frac{IPC_{j}^{shared}}{IPC_{j}^{alone}}}
\end{equation}

\subsubsection{Lock Usage}
To save computation time during training of the actor and the critic ANNs, \methodname uses a heuristic to eliminate \emph{Sched policies} that degrade the mean of the response latency of an application for a given state during the exploration phase (phase where the agent learns the best possible policy by randomly selecting an action). In Section~\ref{sec:motiv_lock}, we highlighted that if the container processes of an LS application use {\em futex} locks on a shared file, then setting the physical CPU allocation degrades the mean response latency. To determine if a serverless application uses futex locks, the state monitor daemon collects the syscalls:sys\_enter\_futex event using the \emph{perf} tool. Subsequently, it sets the Boolean variable $F_{lock}$ in the state of an application.

\subsection{FaaSched}
\label{sec:faasched_design}

\subsubsection{Overview}
We design an \emph{A2C-RL} scheme to set the \emph{Sched policy} of an LS application. The objective of this scheme is to minimize the jitter in the latency of an LS application while ensuring that the $S_{fair}$ metric does not decrease beyond a threshold value equal to $\tau$. To save computation time during training of the actor and the critic ANNs, we eliminate some \emph{Sched policies} for a given state that degrade the mean of the response latency of an application using our novel heuristic function~\cite{state_elimination}. Let us elaborate.

\subsubsection{Features}
\label{sec:rl_features}
The \emph{A2C-RL} scheme uses the following features to decide the \emph{Sched policy} of an LS application $F_{id}$: the contention suffered by the application ($S_{cont}[\,]$), the fairness metric ($S_{fair}$), the $F_{lock}$ attribute, the allocation of CPU resources to the application ($P_{id}$, $A_{id}$) and the allocation of CPU resources to other LS applications ($P_{low}$, $P_{high}$, $A_{other}$). \methodname extracts these features from the state of an application collected by the state monitor daemon process.

\subsubsection{Feature Preprocessing}
To improve the training of the actor and critic models, we need to normalize these features. We normalize the $S_{cont}[\,]$ vector by first linearly converting every element to a number between 0 to 10 using min-max scaling~\cite{minmax_scaling}. Then, we compute the L2 norm~\cite{l21} (square root of sum of squares) of $S_{cont}[\,]$ and use it to divide every element of $S_{cont}[\,]$. Prior work~\cite{l21, l22} has found L2 normalization to be more effective than other methods such as L1 normalization for such vectors. We  then normalize the priority ($P_{id}$), the physical CPU core allocation ($A_{id}$), and the resource allocation of other LS applications ($P_{low}$, $P_{high}$, $A_{other}$) using the min-max normalization technique.

\subsubsection{Sched Policy}
For a given state ($s$) of an LS application, the \emph{A2C-RL} scheme can impose a \emph{Sched policy} from a plausible set of \emph{Sched policies} ($\mathcal{P}$). The \emph{Sched policy} computes a shift in the priority and the physical CPU core allocation of the container processes of the application. Each of the actions is represented in the form of a $\langle \Delta P, \Delta A \rangle$ tuple, where $\Delta P \in \{-\infty, -P_{step}, 0, P_{step}\}$ and $\Delta A \in \{-\infty, -A_{step}, 0, A_{step}\}$. If the value of $\Delta P$ is $-\infty$, then we fallback to the default scheduling policy (SCHED\_OTHER). Similarly, if the value of $\Delta A$ is $-\infty$, then we revoke the physical CPU core allocation.

\subsubsection{Reward function}
In any RL method, the reward function is by far the most important. Note that we aim to minimize jitter in the latency in a set of LS applications while ensuring that there is no monopolization of CPU resources. We also should not ideally degrade the latency of LS applications, whereas a modest degradation in the latency of LD applications is acceptable. Therefore, we considered the following metrics: the magnitude of the $S_{cont}[\,]$ vector post normalization ($=R_{cont}$), and the fairness metric ($S_{fair}$). We formulate the reward function as a set of piecewise yet discontinuous  linear equations that use the $R_{cont}$ metric and the $R_{fair}$ metric as defined in Equation~\ref{equ:reward}. The principles are as follows: higher the fairness $\Rightarrow$ higher the reward and lower the contention $\Rightarrow$ higher the reward.  In the case, an action attempts to oversubscribe the number of cores available in the system or exceed the range of possible values $(P_{min},P_{max})$ of the priority, we punish the agent by providing a negative reward. We set the scheduling policy to SCHED\_OTHER and revoke the physical CPU core allocation (no affinity).

\begin{equation}
\label{equ:reward}
\begin{aligned}
	&&& R = 
		\begin{cases}
		-c & \text{if $A_{used}$} > \# \text{cores} \\
		-c & \text{if $P_{used}$} > \# \text{$P_{max}$} \\
		-c & \text{if $P_{used}$} < \# \text{$P_{min}$} \\
		a\,R_{fair} - b\,R_{cont} &  \text{otherwise} \\
		\end{cases} \\ 
	& where, && R_{fair} =
		\begin{cases}
		S_{fair} & if \,\, S_{fair} > \tau \\
		0 & \text{otherwise}
		\end{cases} \\
	&&& A_{used} = A_{id} + A_{other} + \Delta A \\
	&&& P_{used} = P_{id} + \Delta P \\
	&&& \text{$a$. $b$, and $c$ are positive constants.}
\end{aligned}
\end{equation}

\begin{figure*}[!htb]
	\centering
	\begin{subfigure}[b]{0.48\linewidth}
		\captionsetup{justification=centering}
		\includegraphics[width=\linewidth]{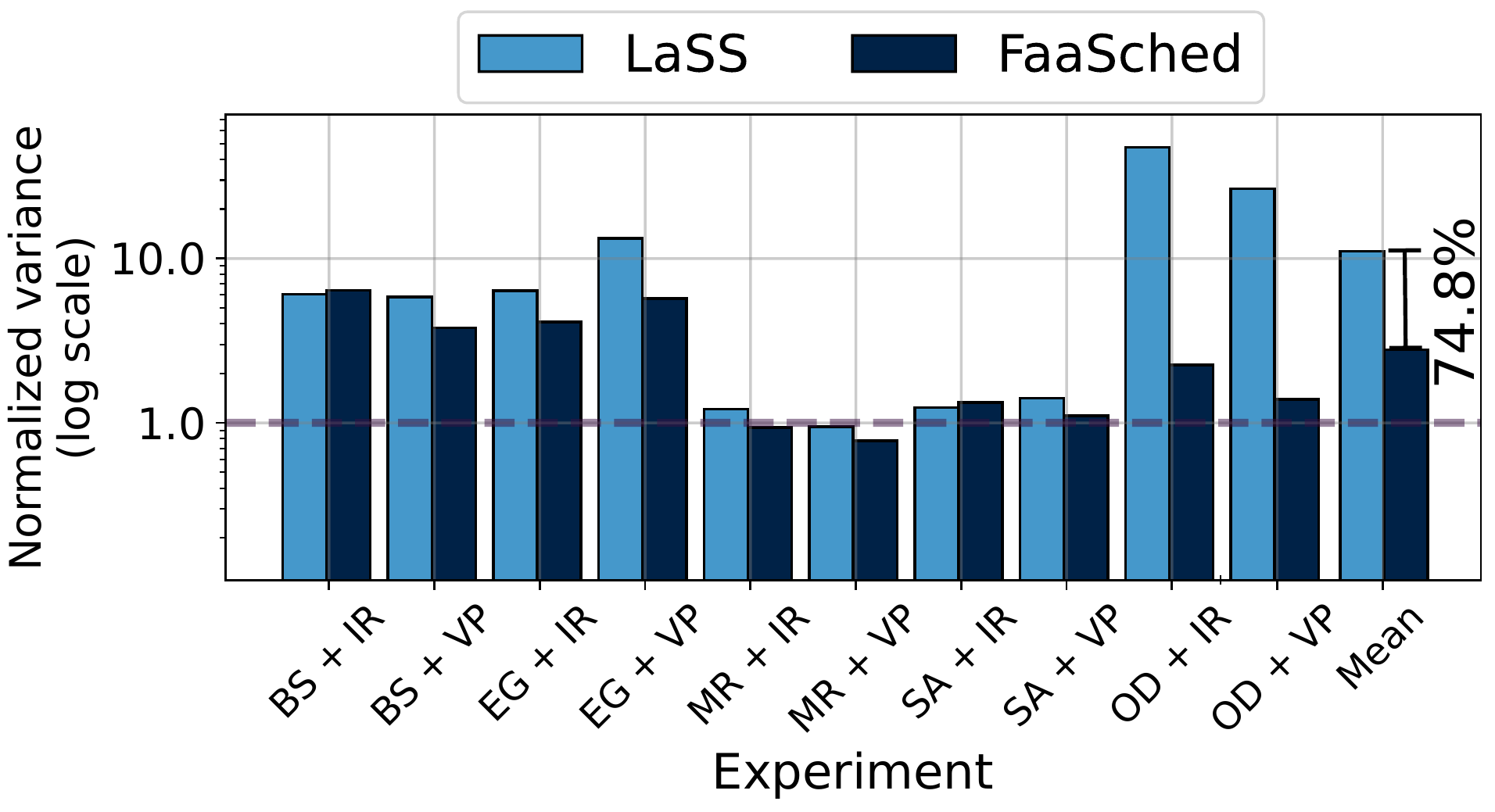}
		\caption{Variance in the response latency}
	\end{subfigure}
	\hfil
	\begin{subfigure}[b]{0.48\linewidth}
		\captionsetup{justification=centering}
		\includegraphics[width=\linewidth]{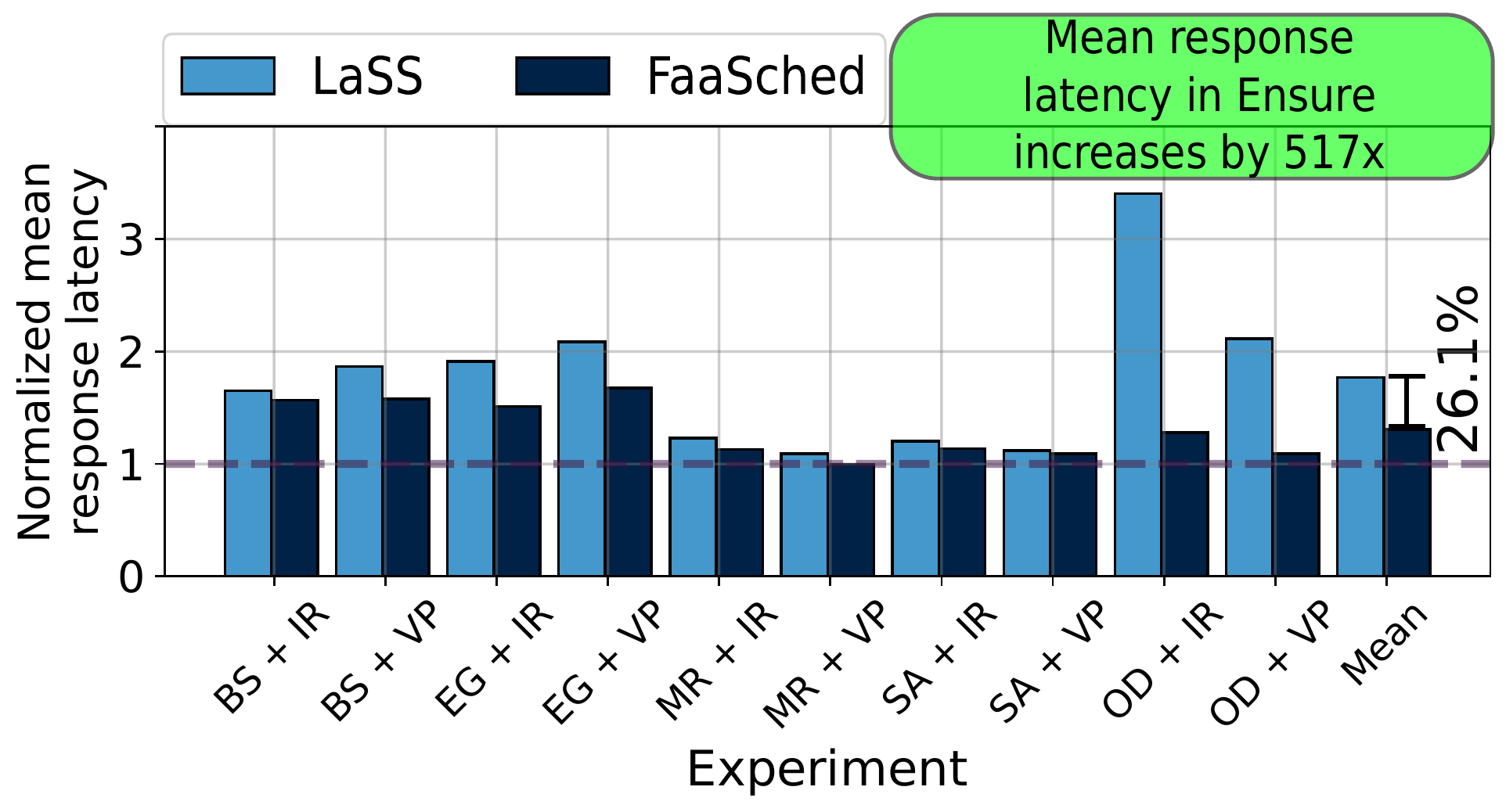}
		\caption{Mean response latency}
	\end{subfigure}
	
	\caption{Comparison of the variance and the mean of the response latency of an LS application when it is executed along with an LD application (normalized to an isolated execution of the LS application).}
	\label{eval:faasched_ls_var}
\end{figure*}

\begin{figure}[!htb]
	\centering
	\captionsetup{justification=centering}
	\includegraphics[width=\linewidth]{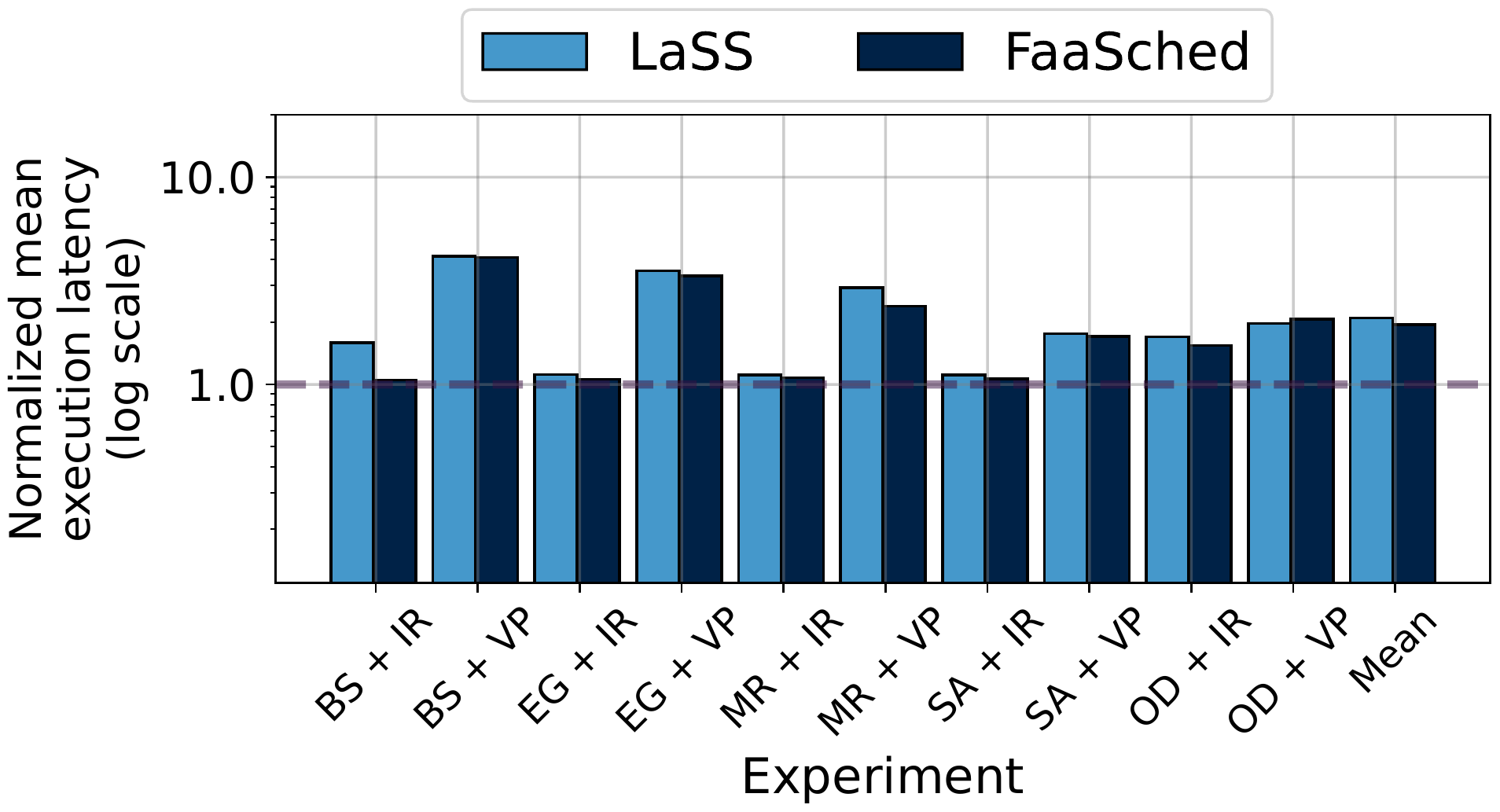}
	\caption{The mean execution latency of an LD application when running along with an LS application (normalized to an isolated execution of the LD application).}
	\label{eval:faasched_ld}
\end{figure}

\subsubsection{Sched Policy Enforcement and Overall Working}
To improve the performance of our RL scheme, we employ the $\epsilon$-greedy algorithm~\cite{eps_greedy} during the exploration phase. In the $\epsilon$-greedy algorithm, the agent selects a random action over the best possible action (till known) for a given state with a probability of $\epsilon$. To save computation time during training of the actor and the critic ANNs~\cite{state_elimination}, we eliminate the policies where $\Delta A \in \{-A_{step}, 0, A_{step}\}$ during the exploration phase if the $F_{lock}$ attribute is set (= 1) in the state of an application. This is because applications employing futex locks should not have a physical CPU affinity (as discussed in Section~\ref{sec:motiv_lock}).

We have used a policy where we do not have offline training (exploration) and online inferencing (exploitation). For us, both the phases are interleaved. This is similar to any architectural prediction scheme such as branch prediction for instance. We start with an exploration:exploitation ratio of 5:1 and then linearly decrease this ratio to 1:100 over a period of 5 hours. Then we run workload combinations that haven't been seen in this time frame -- clean separation between train and test.

\section{Evaluation}
\label{sec:eval}

In this section, we evaluate the jitter using \methodname and compare it against LaSS~\cite{lass} and
Ensure~\cite{ensure}. Recall that we had already discussed the evaluation setup and benchmarks in
Section~\ref{sec:characterization}. We implement these designs on a popular open source serverless framework,
\emph{Apache OpenWhisk v1.0}~\cite{openwhisk}. In Table~\ref{tab:hyperparameters}, we show the values of the
hyperparameters that we use in our experiments (they were found empirically). We shall discuss the robustness of our
hyperparameter choices in Section \ref{sec:eval_hyperparameter}.

\begin{table}
	\centering
	\begin{center}
		\footnotesize
		\caption{Hyperparameters in the \methodname design.}
		\label{tab:hyperparameters}
		\begin{tabular}{|l|p{5.5cm}|p{1cm}|}
			\hline
			\multicolumn{3}{|c|}{\textbf{Hyperparameters}} \\ \hline 
			$\tau$ & The minimum allowed value of the fairness metric ($S_{fair}$). & 0.75 \\ \hline
			$P_{step}$ & A scalar value representing the change in the priority of an application. & 10\\ \hline
			$A_{step}$ & A scalar value representing the change in the physical CPU allocation of an application. & 2\\ \hline
			$\epsilon$ & The probability of taking a random action during the exploration phase & 0.3\\ \hline
			$a$ & The weight of the $R_{fair}$ metric in the reward function. & 1000\\ \hline
			$b$ & The weight of the $R_{cont}$ metric in the reward function. & 100\\ \hline
			$c$ & The penalty value if the agent oversubscribes the total number of cores or provides an priority that 
is not in the acceptable range ([$P_{min}$, $P_{max}$]). & 1000\\ \hline
		\end{tabular}
	\end{center}
\end{table}

To fine tune the hyperparameters and analyze the impact of \methodname on LS and LD applications in detail, we evaluate
\methodname for all possible combinations of a single LS and a single LD application first. In
Section~\ref{eval:scalability}, we evaluate \methodname on a full execution (5 LS + 5 LD applications). \emph{Note: We
use the same values of hyperparameters in all scenarios.}

\begin{table*}
	\centering
	\caption{Detailed statistics of the mean execution latency, the CPU wait time, \emph{nvcs} events, voluntary context switch events and iTLB misses of LS applications.}
	\label{tab:eval_analysis}
	\begin{tabular}{|c|cc|cc|cc|cc|cc|}
		\hline
		\multirow{2}{*}{\textbf{LS Applications}} & \multicolumn{2}{c|}{\textbf{\begin{tabular}[c]{@{}c@{}}Mean
execution\\ latency (msec)\end{tabular}}} & \multicolumn{2}{c|}{\textbf{\begin{tabular}[c]{@{}c@{}}CPU wait time\\
(msec)\end{tabular}}} & \multicolumn{2}{c|}{\textbf{\begin{tabular}[c]{@{}c@{}}Non-voluntary\\ context
switches\end{tabular}}} & \multicolumn{2}{c|}{\textbf{\begin{tabular}[c]{@{}c@{}}Voluntary \\ context
switches\end{tabular}}} & \multicolumn{2}{c|}{\textbf{\#iTLB misses}}              \\ \cline{2-11} 
		& \multicolumn{1}{c|}{\textbf{LaSS}}                         & \textbf{FaaSched}                        &
\multicolumn{1}{c|}{\textbf{LaSS}}                    & \textbf{FaaSched}                    &
\multicolumn{1}{c|}{\textbf{LaSS}}                         & \textbf{FaaSched}                         &
\multicolumn{1}{c|}{\textbf{LaSS}}                        & \textbf{FaaSched}                       &
\multicolumn{1}{c|}{\textbf{LaSS}} & \textbf{FaaSched} \\ \hline
		Binary Scanner (BS)                       & \multicolumn{1}{c|}{271.97}                                & 238.36                                   & \multicolumn{1}{c|}{1081}                             & 0.008                                & \multicolumn{1}{c|}{3.8 K}                                 & 0.03 K                                    & \multicolumn{1}{c|}{8.4 K}                                & 8.4 K                                   & \multicolumn{1}{c|}{11.8 M}        & 8.3 M             \\ \hline
		Email Generator (EG)                      & \multicolumn{1}{c|}{294.07}                                & 275.76                                   & \multicolumn{1}{c|}{749.4}                            & 2.7                                  & \multicolumn{1}{c|}{5 K}                                   & 0.13 K                                    & \multicolumn{1}{c|}{1.9 K}                                & 2 K                                     & \multicolumn{1}{c|}{202.4 M}       & 152.1 M           \\ \hline
		Markdown Renderer (MR)                    & \multicolumn{1}{c|}{109.5}                                 & 102.8                                    & \multicolumn{1}{c|}{643}                              & 637.3                                & \multicolumn{1}{c|}{3.8 K}                                 & 0.05 K                                    & \multicolumn{1}{c|}{2.8 K}                                & 2.6 K                                   & \multicolumn{1}{c|}{352.9 M}       & 247.5 M           \\ \hline
		Stock Analyzer (SA)                       & \multicolumn{1}{c|}{397.37}                                & 375.34                                   & \multicolumn{1}{c|}{1341}                             & 1210.7                               & \multicolumn{1}{c|}{30 K}                                  & 0.1 K                                     & \multicolumn{1}{c|}{0.9 K}                                & 0.8 K                                   & \multicolumn{1}{c|}{207.7 M}       & 132.9 M           \\ \hline
		Object Detector (OD)                      & \multicolumn{1}{c|}{230.82}                                & 192.85                                   & \multicolumn{1}{c|}{45679}                            & 74.7                                 & \multicolumn{1}{c|}{100 K}                                 & 0.9 K                                     & \multicolumn{1}{c|}{42.6 K}                               & 22.2 K                                  & \multicolumn{1}{c|}{69 M}          & 60.2 M            \\ \hline
	\end{tabular}
\end{table*}

\begin{figure*}[!htb]
	\centering
	\begin{subfigure}[b]{0.45\linewidth}
		\captionsetup{justification=centering}
		\includegraphics[width=0.9\linewidth]{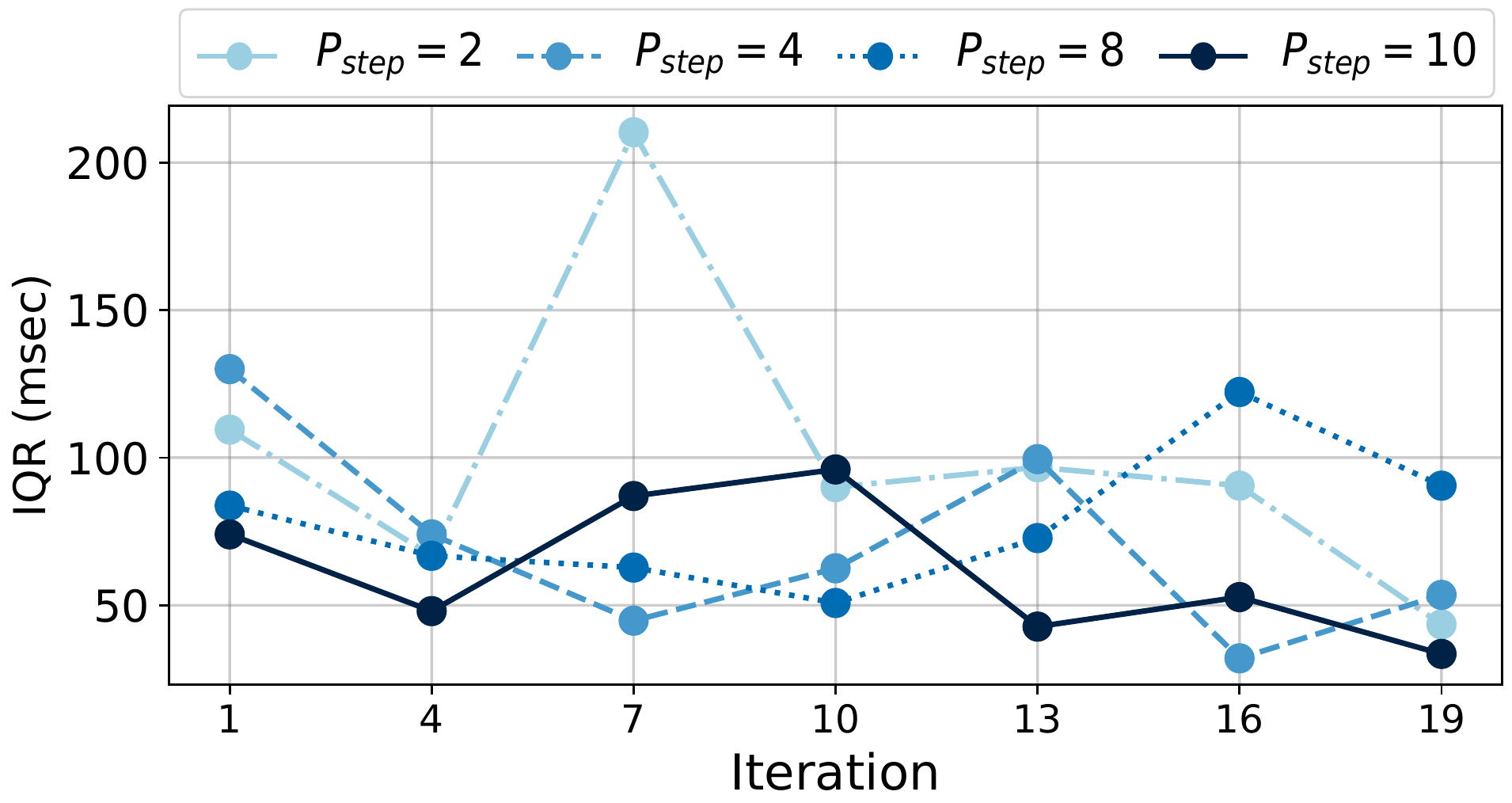}
		\caption{The IQR value with varying the $P_{step}$ value.}
	\end{subfigure}
	\hfil
	\begin{subfigure}[b]{0.45\linewidth}
		\captionsetup{justification=centering}
		\includegraphics[width=0.9\linewidth]{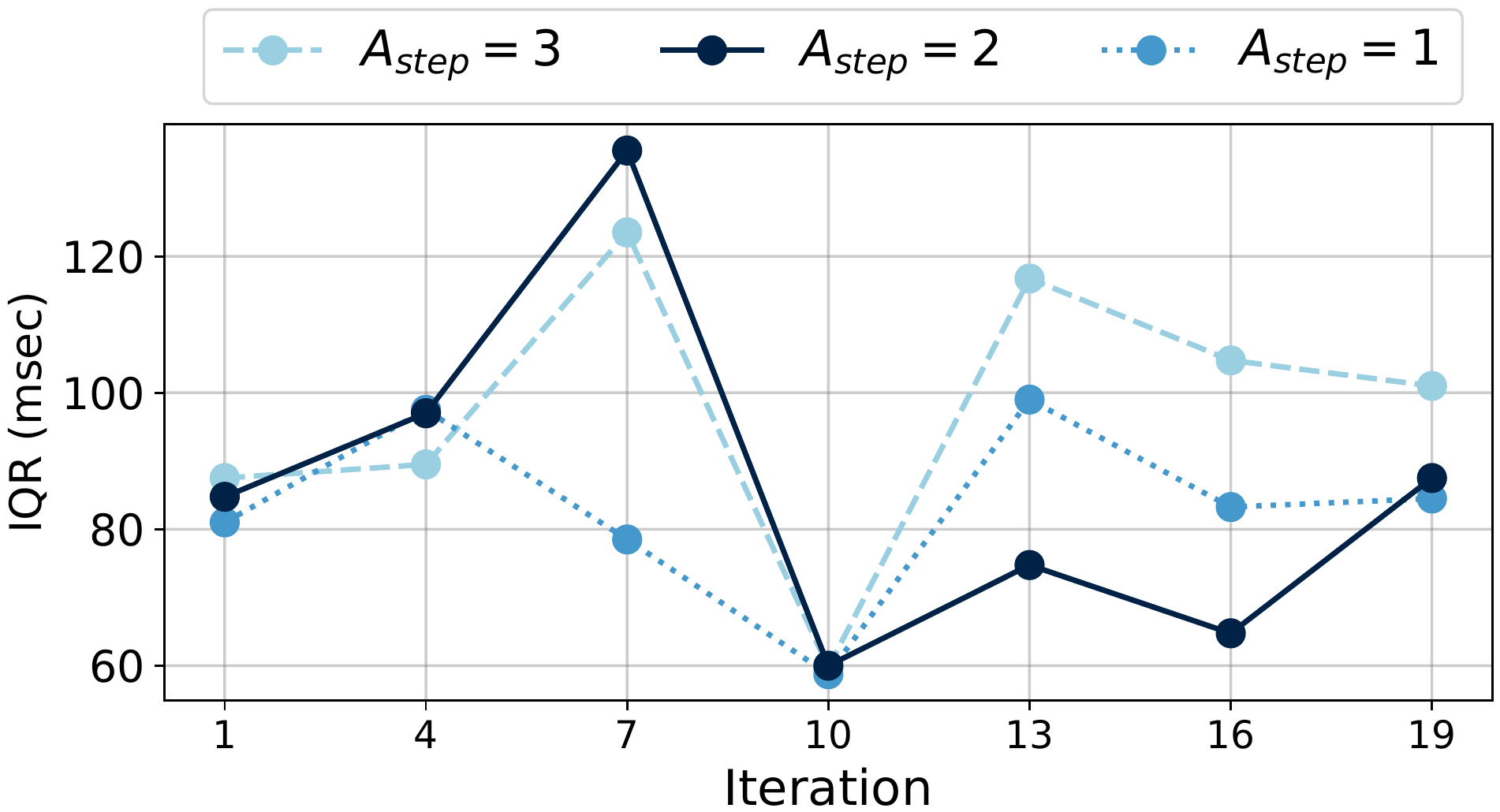}
		\caption{The IQR value with varying the $A_{step}$ value.}
	\end{subfigure}	
    \caption{The IQR values for different $P_{step}$ and $A_{step}$ combinations (EG+VP combination)}
	\label{eval:sensitivity}
\end{figure*}

\begin{figure*}[!htb]
	\centering
	\begin{subfigure}[b]{0.32\linewidth}
		\captionsetup{justification=centering}
		\includegraphics[width=\linewidth]{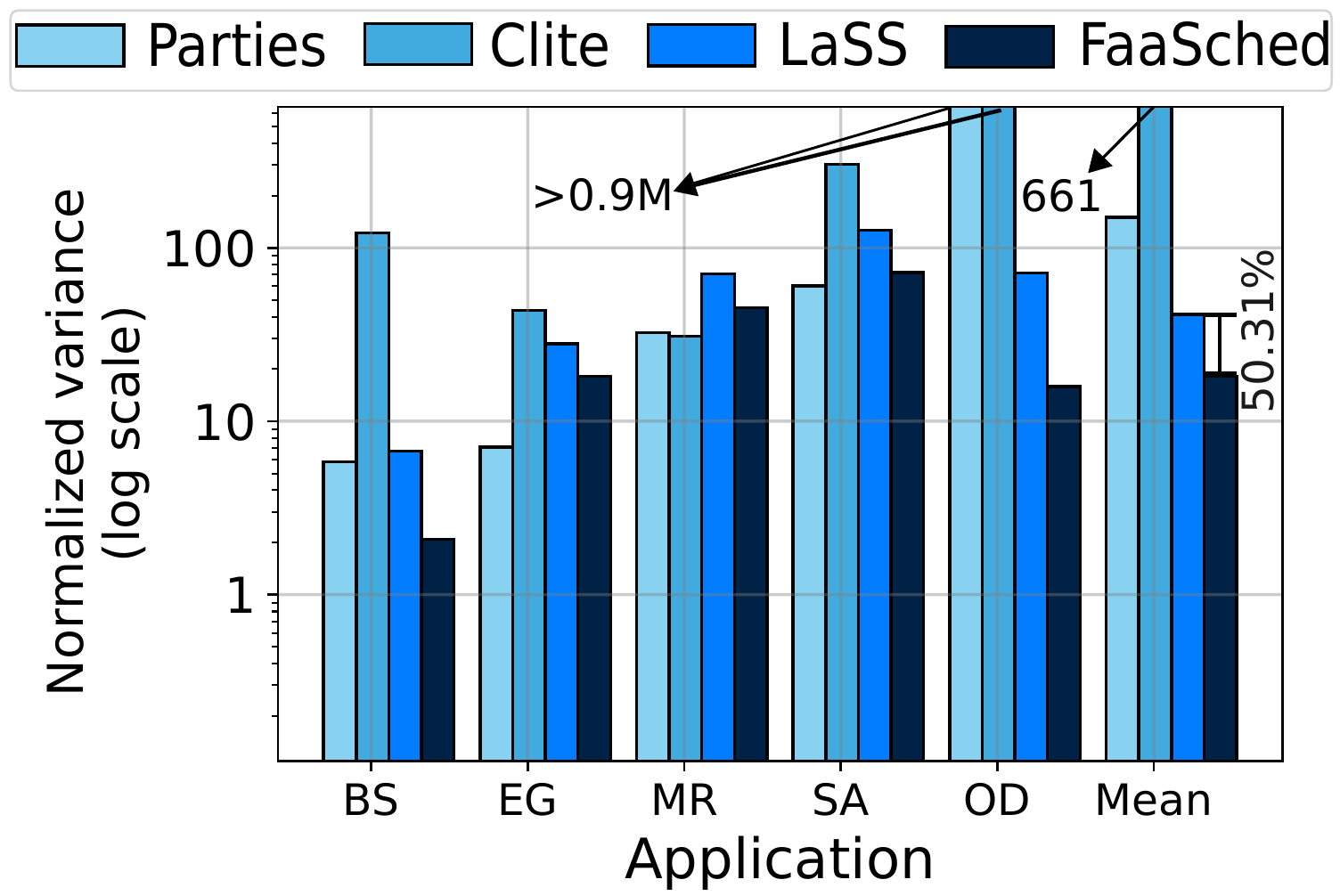}
		\caption{Variance in the response latency of LS applications.}
	\end{subfigure}
	\hfil
	\begin{subfigure}[b]{0.32\linewidth}
		\captionsetup{justification=centering}
		\includegraphics[width=\linewidth]{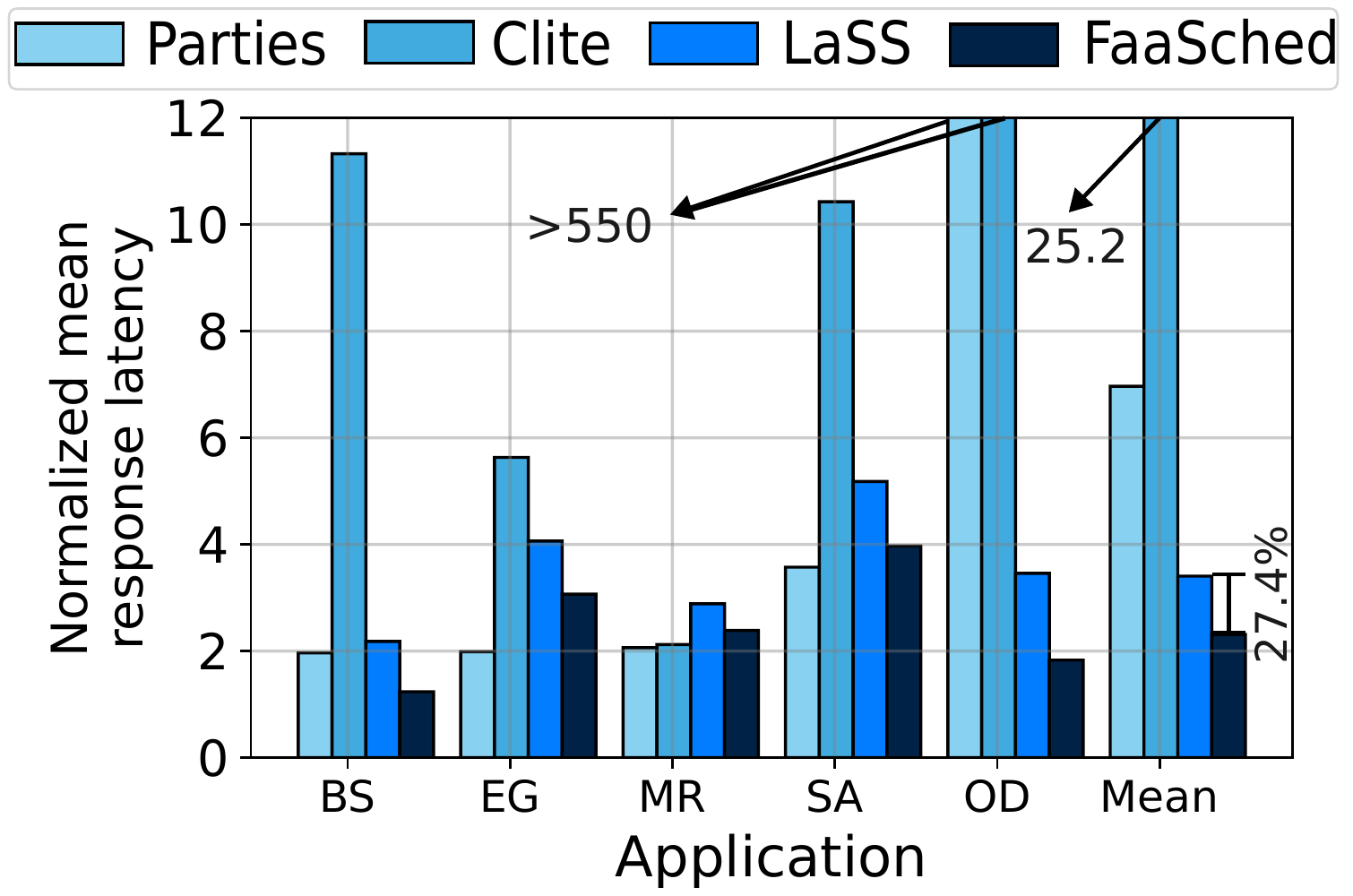}
		\caption{Mean response latency of LS applications.}
	\end{subfigure}
	\hfil
	\begin{subfigure}[b]{0.32\linewidth}
		\captionsetup{justification=centering}
		\includegraphics[width=\linewidth]{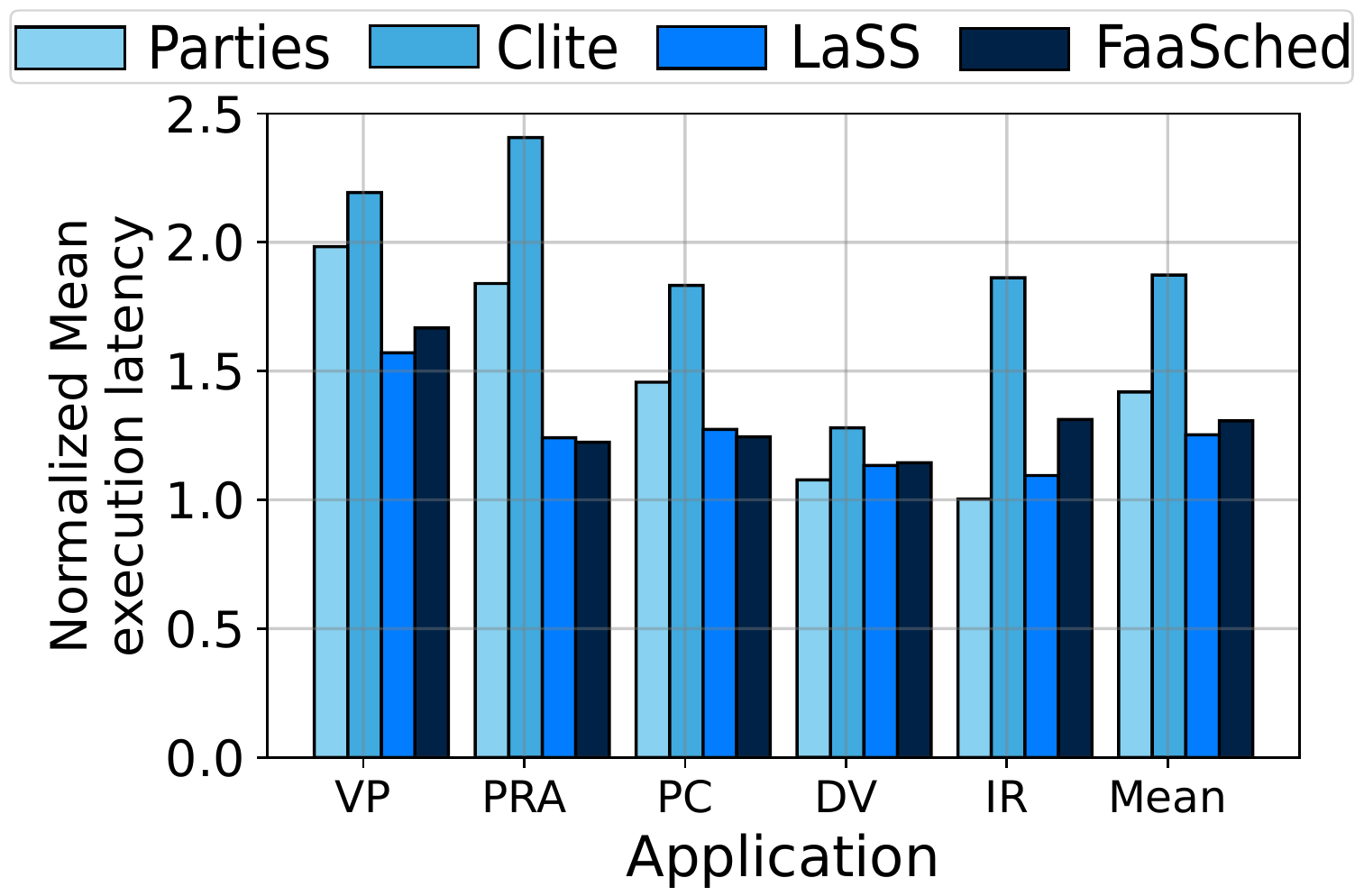}
		\caption{Mean execution latency of LD applications.}
	\end{subfigure}	
    \caption{Statistics related to the execution latency and the response latency of applications when all LS
applications are running along with all LD applications (normalized to an isolated execution).}
	\label{eval:faasched_all}
\end{figure*}

\subsection{Evaluating Jitter and Fairness}

\subsubsection{10 {\em Stressed} Combinations}
To evaluate the jitter in the latency, we measure the variance and the mean of the response latency (time between
delivering a request and receiving the response) of an LS application. The performance is defined as the
reciprocal of the mean response time.  We choose two LD applications such that the system is maximally
stressed: IR (highest arrival rate) and VP (highest service time). Figure~\ref{eval:faasched_ls_var} shows a
comparison between \methodname and LaSS~\cite{lass}.

We observe that \methodname improves the variance and the mean of the response latency of LS applications by 74.8\% and
26.1\% (resp.), on an average (normalized to LaSS). This is happening because of the superior simulations
provided by our RL algorithm. As a
result, the CPU wait time, \emph{\#nvcs} events, and {\em \#iTLB\_misses} reduce by up to 99.5\%, 99.7\%, and 12\%,
respectively. Effectively, our RL algorithm is eliminating all non-voluntary context switches because the associated
interrupts and kernel events are routed to other cores that run LD applications. The wait time of the LS application
(total time spent in the run queues) also gets reduced to a near-nil value for the same reason; LD applications are
pre-empted because of their lower priority. We observe that Ensure~\cite{ensure} significantly underperforms because it
naively allocates and deallocates the container processes of an application; thus, multiple requests suffer
from a cold start: roughly up to 32\% of requests. In addition to the scheduling cost, it takes up to 3197 msec to spawn
a container. Therefore we observe a mean 517$\times$ increase in the response latency of an application.

For the mean execution latency of the two LD applications -- IR and VP --
refer to Figure~\ref{eval:faasched_ld}. The mean execution latency
degrades by up to 5.14\% (as compared to LaSS), which is minimal. The degradation
is mostly accounted for by one combination (OD+VP).
This is on expected lines; our main aim was to trade-off the latency
of LD applications with the jitter in LS applications. 

\subsubsection{All 25 combinations: $LS \times LD$}
Let us now consider all pairwise combinations of LS and LD workloads. The variance and mean of the response time
for LS applications reduce by 77.12\% and 18.76\% (resp.), on an average, as compared to LaSS. 
For the LD applications, we observe that
\methodname increases the mean execution latency by up to 12.11\%.

\subsection{Analyzing Hyperparameter Sensitivity}
\label{sec:eval_hyperparameter}
As mentioned in Section~\ref{sec:motivation}, we identify two knobs to set the Sched policy of an application: the
priority and the physical CPU core allocation. \methodname changes the aforementioned knobs by a fixed step size
($P_{step}$ and $A_{step}$). We chose a random pair namely (EG+VP) and evaluated the IQR value of
the execution latency of the LS application, EG. Similar trends held for other pairs, other than the ones that had OD.
In Figure~\ref{eval:sensitivity}, we can see that the
best values for $P_{step}$ and
$A_{step}$ are 10 and 2, respectively. 

\subsection{Evaluating Scalability}
\label{eval:scalability}
To evaluate the scalability of \methodname, we measure the variance and the mean of the response latency while executing
all five LS applications along with all five LD applications simultaneously. The experiments were performed on a machine
with an Intel Xeon 6226R
CPU (2.90GHz, 16 cores) with 256GB RAM. Figure~\ref{eval:faasched_all} shows a comparison between Parties~\cite{parties}, Clite~\cite{clite}, \methodname and
LaSS~\cite{lass}. We observe that \methodname improves the variance and the mean of the response latency of LS
applications by 50.31\% and 27.4\% (resp.), on an average, as compared to LaSS.  We observe that the CPU wait time,
\emph{\#nvcs} events and {\em \#iTLB\_misses} in LS applications reduces by up to 99.9\%, 99.7\%, and 36\% as compared to
LaSS (as shown in Table~\ref{tab:eval_analysis}). We also observe that the mean execution latency of LD applications
degrades by up to 19.88\%, as compared to LaSS (as shown in Figure~\ref{eval:faasched_all}).

However, in some LD applications, we observe that the execution latency decreases by up to 2.23\%  (see
Figure~\ref{eval:faasched_all}). This is because \methodname allocates dedicated CPU cores to LS applications, thus
providing resource isolation between LS applications and LD applications. As a result, the \emph{\#iTLB\_Misses}
suffered by an LD application reduces by up to 5\%. Similar behavior has been seen in ~\cite{core-partition}.

Parties~\cite{parties} improves the mean response latency and variance of the serverless applications (except the binary scanner and object detector) compared to LaSS. This is because it has set the best resource configuration for the applications. However, in Figure~\ref{eval:faasched_all}, we show that the average execution latency of LD applications degrades by up to 48.22\%. \methodname explicitly rewards or punishes the RL model based on the degradation of LD applications, thus ensuring a greater degree of fairness.

\section{Related Work}
\label{sec:related_work}

\subsection{Resource Scheduling in Microservice}
Prior work~\cite{parties, sinan, clite} utilize resource partitioning techniques to ensure that latency-sensitive jobs meet their respective QoS guarantees. Parties~\cite{parties} alters one of the allocated resources (CPU core affinity, LLC way partition, memory capacity, and CPU core frequency) of an LS application by a fixed size at a time depending on the performance of the application. While Clite~\cite{parties} uses Bayesian optimization to find an optimal resource configuration (CPU core affinity, memory bandwidth, and LLC way partition) for LS applications. In Section~\ref{sec:characterization}, we found that the CPU wait time of an application significantly impacts the response latency of an application in the serverless computing paradigm. Moreover, each application consists of three processes (as discussed in Section~\ref{sec:background}). Existing resource partitioning techniques do not take any measures to minimize the CPU wait time of an application during colocation.

\subsection{Resource Scheduling in Serverless Computing}
Prior work~\cite{openwhisk_jitter,catalyzer,nightcore,lass,interference} focused on improving the latency of an application but did not consider the jitter in the latency in a multi-tenant setup. To improve the latency, prior work~\cite{lass,ensure,placement2} focused on designing efficient resource scheduling schemes. They have highlighted that the degradation in the latency (w.r.t. native execution) is due to sharing limited resources on a single host machine~\cite{interference}.

Kaffes et. al.~\cite{placement2} proposed a CPU core partitioning technique that assigns a dedicated CPU core to each request, thus minimizing resource contention. Nevertheless, when the load factor is high, allocating a dedicated core to each request has an effect on the waiting time of subsequent requests. Instead of using the core partitioning technique, Ensure~\cite{ensure} dynamically increases the CPU time of a sandbox process assigned to an application to mitigate resource contention. This scheme does not consider resource contention in other LS applications while taking an action -- this has adverse implications on the overall fairness. In our work, we set the \emph{Sched policy} of the container processes of an application to minimize jitter in latency by capturing the complete state of the system (contention suffered by all the applications running on the system).

Instead of using heuristics, LaSS~\cite{lass} uses a queuing theory-based model to
determine the ideal number of container processes (that must be spawned for an application) to limit the waiting time. Furthermore, it maintains a few container processes of an application in a standby mode to serve future requests. It does not explicitly consider the effects of colocation. 

There is prior work that looks at cluster-level scheduling~\cite{placement1, placement3, openwhisk_jitter, real_faas}, which is orthogonal to our work. We only look at scheduling decisions inside a single physical machine. This can be coupled with efficient cluster-level scheduling schemes. The most important point to note is that prior work hasn't looked at ensuring determinism in terms of execution time as we do.

\subsection{RL-based Resource Scheduling}

Grid computing~\cite{grid_rl}, the Android OS~\cite{rl_cpu1} and cloud computing~\cite{rl_cpu2} are just a few of the many fields that use RL-based schemes to schedule resources across applications in order to improve performance. In a cloud computing system, applications typically exhibit time-varying resource usage patterns. To learn the temporal resource usage pattern of applications, Mondal et. al.~\cite{rl_cpu2} and VCONF~\cite{vconf} employ an RL scheme to compute the allocation of memory and CPU resources. They use the following features in their state representation: resource availability, resource allocation, and resource utilization of the cluster. To evaluate an action, they use the following metrics: the resource contention, the wait time (time spent by a request waiting for execution), the throughput, and the total number of service level agreement violations.

In contrast to prior work, \methodname uses the contention suffered by an application along with the resource allocation of applications in the state representation supplemented with some microarchitectural counter values such as \#iTLB\_misses. The reward function explicitly includes the \emph{fairness} metric (as discussed in Section~\ref{sec:design}) along with resource contention to restrict applications from monopolizing resources.
\section{Conclusion}
\label{sec:conclusion}

In this paper, we introduce a new paradigm that {\em trades off} jitter in LS applications with the performance of LD applications. To the best of our knowledge, this has not been done before in the area of serverless applications. To do so, we had to perform a detailed characterization study and identify the metrics of interest such as iTLB based events, \#nvcs events and the CPU wait time. Designing an RL-based scheme for our two knobs was not easy because there are many different reward functions and choosing the best one that restricts monopolization of CPU resources and oversubscription of cores was non-trivial. Even properly representing the state of the system that captures aspects of our interest required a great deal of thinking. Finally, we were able to show that our design is scalable; for 10 applications (5 LS + 5 LD), we showed that we reduce the variance of LS applications by 50.31\%, decrease their response time by 27.4\%, and limit the execution time degradation of LD applications to 19.88\% compared to LaSS~\cite{lass}. \methodname increases the response time of LS applications by 150\% over the isolated execution as compared to 255\% for LaSS.

\newpage
\bibliographystyle{IEEEtranS}
\bibliography{references}

\end{document}